\documentclass[runningheads]{llncs}

\usepackage{macro}
\usepackage{pifont}
\newcommand{\xmark}{\ding{55}}%
\usepackage{fsa}
\usepackage{shortcuts}

\begin{document}
\title{Twinning automata and regular expressions for string static analysis}
\titlerunning{Twinning automata and regular expressions for string static analysis}

\author{Luca Negrini \inst{1,2} \and Vincenzo Arceri \inst{1} \and Pietro Ferrara \inst{1} \and Agostino Cortesi\inst{1}}


\institute{
	Ca' Foscari University of Venice, Venice, Italy \\
	\email{vincenzo.arceri@unive.it $\;$ pietro.ferrara@unive.it $\;$ cortesi@unive.it}\and
	JuliaSoft S.r.l., Verona, Italy\\
	\email{luca.negrini@juliasoft.com}
}

\maketitle              
\begin{abstract}

In this paper we formalize and prove the soundness of $\tarsis$, a new abstract domain based on the abstract interpretation theory that approximates string values through finite state automata. The main novelty of $\tarsis$ is that it works over an alphabet of strings instead of single characters. On the one hand, such approach requires a more complex and refined definition of the widening operator, and the abstract semantics of string operators. On the other hand, it is in position to obtain strictly more precise results than than state-of-the-art approaches. We implemented a prototype of $\tarsis$, and we applied it on some case studies taken from some of the most popular Java libraries manipulating string values. The experimental results confirm that $\tarsis$ is in position to obtain strictly more precise results than existing analyses.

\keywords{String analysis  \and Static analysis \and Abstract interpretation.}
\end{abstract}

\section{Introduction}\label{sec:intro}

Strings play a key role in any programming language due to the many and different ways in which they are used, for instance to dynamically access object properties, to hide the program code by using string-to-code statements and reflection, or to manipulate data-interchange formats, such as JSON, just to name a few. Despite the great effort spent in reasoning about strings, static analysis often failed to manage programs that heavily manipulate strings, mainly due to the inaccuracy of the results and the prohibitive amount of resources (time, space) required  to retrieve useful information on strings. One the one hand, finite height string abstractions~\cite{costantini2015} are computable in a reasonable time, but precision is suddenly lost when using advanced string manipulation. On the other hand, more sophisticated abstractions (e.g., the ones reported in~\cite{arceri2019-fa,mstring}) compute precise results but they require a huge, and sometimes unrealistic, computational cost, making such code intractable for these abstractions. A good representative of such abstractions is the finite state automata domain~\cite{arceri2019-fa}. Over-approximating strings into finite state automata has shown to increase string analysis accuracy in many scenarios, but it does not scale up to real world programs dealing with statically unknown inputs and long text manipulations.

The problem of statically analyzing strings has been already tackled in different contexts in the literature~\cite{mstring,arceri2019-fa,park2016,sas2003,madsen2014,abdulla2014,costantini2015}. The original finite state automata abstract domain has been defined in~\cite{arceri2019-fa} in the context of dynamic languages, providing an automata-based abstract semantics for common ECMAScript string operations. The same abstract domain has been integrated also for defining a sound-by-construction analysis for string-to-code statements~\cite{arceri2020}. The authors of~\cite{otherautomata} provided an automata abstraction merged with interval abstractions for analyzing JavaScript arrays and objects. In~\cite{sas2003}, the authors propose static analysis of Java strings based on the abstraction of the control-flow graph as a context-free grammar. Regular strings~\cite{regstrings} is an abstraction of the finite state automata domain and approximates strings as a strict subset of regular expressions. Even if it is not tackled the problem of analyzing strings, in~\cite{midtgaard2016} is proposed a lattice-based generalization of regular expressions, showing a regular expressions-based domain parametric from a lattice of reference. Finally, automata have been also involved in model checking in order to tackle the well-known problem of state space explosion~\cite{bouajjani2004,bouajjani2006}.

\vskip10pt

In this paper we introduce $\tarsis$, a new abstract domain for string values based on finite state automata (FSA). Standard FSA has been shown to provide precise abstractions of string values when all the components of such strings are known, but with high computational cost. Instead of considering standard finite automata built over an alphabet of single characters, $\tarsis$ considers automata that are built over an alphabet of strings. The alphabet comprises a special value to represent statically unknown strings. This avoids the creation of self-loops with any possible characters as input, which otherwise would significantly degrade performance. We define the abstract semantics of mainstream string operations, namely \code{substring}, \code{length}, \code{indexOf}, \code{replace}, \code{concat} and \code{contains}, either defined directly on the automaton or on its corresponding equivalent regular expression. Soundness proofs are provided for a subset of the operations.

$\tarsis$ has been implemented into a prototypical static analyzer supporting a subset of Java. By comparing $\tarsis$ with other cutting-edge domains for string analysis, results show that (i) when applied to simple code that causes a precision loss in simpler domains, $\tarsis$ correctly approximate string values within a comparable execution time, (ii) on code that makes the standard automata domain unusable due to the complexity of the analysis, $\tarsis$ is in position to perform in a limited amount of time, making it a viable domain for complex and real codebases, and (iii) $\tarsis$ is able to precisely abstract complex string operations that have not been addressed by state-of-the-art domains.

The rest of the paper is structured as follows. Sect.~\ref{sec:motivating} introduces a motivating example. Sect.~\ref{sec:bg} defines the mathematical notation used throughout the paper. Sect.~\ref{sect:domain} formalizes $\tarsis$ and its abstract semantics. Sect.~\ref{sec:experiments} reports experimental results and comparison with other domains, while Sect.~\ref{sec:conclusion} concludes. Selected proofs can be found in  Appendix~\ref{sect:proofs}.

\section{Motivating example}
\label{sec:motivating}


Consider the code of Fig.~\ref{code:countmatches}, that counts the occurrences of string  {\tt sub} into string {\tt str}. This code is (a simplification of) the \textit{Apache commons-lang} library method \code{StringUtils.countMatches} \footnote{{\tt \url{https://commons.apache.org/proper/commons-lang/}}}, one of the most popular Java libraries providing extra functionalities over the core classes of the Java lang library (that contains class {\tt String} as well). Proving properties about the value of \code{count} after the loop is particularly challenging, since it requires to correctly model a set of string operations (namely, {\tt length}, {\tt contains}, {\tt indexOf}, and {\tt substring}) and their interaction. State-of-the-art string analyses fail to model precisely most of such operations, since their abstraction of string values is not rigorous enough to deal with such situations. Such loss of precision usually leads to fail to prove string-based properties (also on non-string values) in real-world software, such as the numerical bounds of the value returned by method {\tt countMatches} when applied to some string values. 

The goal of this paper is to provide abstract interpretation-based static analysis, in order to deal with complex and nested string manipulations similar to the one reported in Fig.~\ref{code:countmatches}. As we will discuss in Sect.~\ref{sec:experiments}, $\tarsis$ models (among the others) all string operations used in \code{countMatches}, 
and it is precise enough to infer, given the abstractions of \code{str} and \code{sub}, the precise range of values that \code{count} might have at the end of the method.

\begin{figure}[t]
	\begin{CenteredBox}
		\begin{lstlisting}
		int countMatches(String str, String sub) {
		int count = 0;
		int len = sub.length();
		while (str.contains(sub)) { 
		int idx = str.indexOf(sub);
		count = count + 1;
		int start = idx + len;
		int end = str.length();
		str = str.substring(start, end);
		}
		return count;
		}
		\end{lstlisting}
	\end{CenteredBox}
	\caption{A program that counts the occurrences of a string into another one}
	\label{code:countmatches}
	\vskip-15pt
\end{figure}

\section{Preliminaries}\label{sec:bg}

\noindent\textbf{Mathematical notation.} $\;$ Given a set $S$, $S^*$ is the set of all finite sequences of elements of $S$. If $s = s_0\dots s_{n}\in S^*$, $s_i$ is the $i$-th element of $s$, $|s| = n + 1$ is its length, and $s[x / y]$ is the sequence obtained replacing all occurrences of $x$ in $s$ with $y$. When $s'$ is a subsequence of $s$, we write $s' \sub s$. We denote by $s^n, n \ge 0$ the $n$-times repetition of the string $s$. 
Given two sets $S$ and $T$, $\wp(S)$ is the powerset of $S$, $S \smallsetminus T$ is the set difference, $S \subset T$ is the strict inclusion relation between $S$ and $T$, $S \subseteq T$ is the inclusion relation between $S$ and $T$, and $S \times T$ is the Cartesian product between $S$ and $T$. 

\noindent\textbf{Ordered structures.} $\;$
A set $L$ with a partial ordering relation $\leq \subseteq L \times L$ is a poset, denoted by $\tuple{L,\leq}$. A poset $\tuple{L,\leq, \vee , \wedge}$, where $\vee$ and $\wedge$ are respectively the least upper bound (lub) and greatest lower bound (glb) operators of $L$, is a lattice if $\forall x,y \in L \st x \vee y$ and $x \wedge y$ belong to $L$. It is also complete if  $\forall X \subseteq L$ we have that $\bigvee X,\bigwedge X\in L$. 
A complete lattice $L$, with ordering $\leq$, lub $\vee$, glb $\wedge$, top element $\top$, and bottom element $\bot$ is denoted by $\tuple{L,\leq,\vee,\wedge,\top,\bot}$. 

\noindent\textbf{Abstract interpretation.} $\;$
Abstract interpretation~\cite{cc77,cc79} is a theoretical framework for sound reasoning about semantic properties of a program, establishing a correspondence between the concrete semantics of a program and an approximation of it, called abstract semantics. Let $C$ and $A$ be complete lattices, a pair of monotone functions $\alpha: C \rightarrow A$ and $\gamma: A \rightarrow C$ forms a \emph{Galois Connection} (GC) between $C$ and $A$ if $\forall x \in C, \forall y \in A : \alpha(x) \leq_A y \Leftrightarrow x \leq_C \gamma(y)$. We denote a GC as $C \galois{\alpha}{\gamma} A$.
Given $C\galois{\alpha}{\gamma}A$, a concrete function $f: C \rightarrow C$ is, in general, not computable. Hence, a function $f^\sharp : A \rightarrow A$ that must \textit{correctly} approximate the function $f$ is needed. If so, we say that  the function $f^\sharp$ is \textit{sound}. Given $C \galois{\alpha}{\gamma} A$ and a concrete function $f : C \rightarrow C$, an abstract function $f^\sharp : A \rightarrow A$ is sound w.r.t. $f$ if $\forall c \in C.\:\alpha(f(c)) \leq_A f^\sharp(\alpha(c))$.
Completeness~\cite{giacobazzi2000} can be obtained by enforcing the equality of the soundness condition and it is called \textit{backward completeness}. Given $C \galois{\alpha}{\gamma} A$, a concrete function $f : C \rightarrow C$ and an abstract function $f^\sharp : A \rightarrow A$, $f^\sharp$ is backward complete w.r.t. $f$ if $\forall c \in C.\:\alpha(f(c)) = f^\sharp(\alpha(c))$.

\noindent\textbf{Finite state automata and regular expression notation.} $\;$ We follow the notation reported in~\cite{arceri2019-fa} for introducing finite state automata. A finite state automaton (FA) is a tuple $\aut = \tuple{Q, \Sigma, \delta, q_0, F}$, where $Q$ is a finite set of states, $q0 \in Q$ is the initial state, $\Sigma$ is a finite alphabet of symbols, $\delta \subseteq Q \times \Sigma \times Q$ is the transition relation and $F \subseteq Q$ is the set of final states. If $\delta : Q \times \Sigma \rightarrow Q$ is a function then $\aut$ is called deterministic finite state automaton. The set of all the FAs is $\fa$. If  $\lang \subseteq \Sigma^*$  is recognized by an FA, we say that $\lang$ is a regular language.
Given $\aut\in\fa$, $\lang(\aut)$ is the language accepted by $\aut$. From the Myhill-Nerode theorem, for each regular language uniquely exists a minimum FA (w.r.t. the number of states) recognizing the language. Given a regular language $\lang$, $\minimize(\aut)$ is the minimum FA $\aut$ s.t. $\lang = \lang(\aut)$. Abusing notation, given a language $\lang$, $\minimize(\lang)$ is the minimal FA recognizing $\lang$. 
We denote as $\paths{\aut} \in \wp(\delta^*)$ the set of sequences of transitions corresponding to all the possible paths from the initial state $q_0$ to a final state $q_n \in F$. Given $\pi \in \paths{\aut}$, $|\pi|$ is its length, meaning the sum of the lengths of the symbols that appear on the transitions composing the path. Furthermore, $\minpath{\aut} \in \paths{\aut}$ and $\maxpath{\aut} \in \paths{\aut}$ are the paths of minimum and maximum length, respectively. Given $\pi = t_0\dots t_n \in \mathsf{paths}(\aut)$, $\sigma_{\pi_{i}}$ is the symbol read by the transition $t_i$, $i \in [0,n]$, and $\sigma_\pi =\sigma_{\pi_0}\dots\sigma_{\pi_n}$ is the string recognized by such path. Predicate $\hascycle{\aut}$ holds if and only if the given automaton contains a loop. Throughout the paper, it could be more convenient to refer to a finite state automaton by its regular expression (regex for short), being equivalent. Given two regexes $\re_1$ and $\re_2$, $\re_1 \ || \ \re_2$ is the disjunction between $\re_1$ and $\re_2$, $\re_1\re_2$ is the concatenation of $\re_1$ with $\re_2$, $(\re_1)^*$ is the Kleene-closure of $\re_1$.


\noindent\textbf{The finite state automata abstract domain.} $\;$ Here, we report the necessary notions about the finite state automata abstract domain presented in~\cite{arceri2019-fa}, over-approximating string properties as the minimum deterministic finite state automaton recognizing them. 
Given an alphabet $\Sigma$, the finite state automata domain is defined as $\latticefa$, where $\fa$ is the quotient set of $\DFA$ \wrt the equivalence relation induced by language equality, $\leqfa$ is the partial order induced by language inclusion, $\lubfa$ and $\glbfa$ are the lub and the glb, respectively. The minimum is $\minimize(\varnothing)$, that is, the automaton recognizing the empty language and the maximum is $\minimize(\Sigma^*)$, that is, the automaton recognizing any possible string over $\Sigma$. We abuse notation by representing equivalence classes in $\fa$ by one of its automaton (usually the minimum), \ie when we write $\aut\in\fa$ we mean $[\aut]_{\equiv}$. Since $\fa$ does not satisfy the Ascending Chain Condition (ACC), \ie it contains infinite ascending chains, it is equipped with the parametric widening $\widfa$. The latter is defined in terms of a state equivalence relation merging states that recognize the same language, up to a fixed length $n \in \nats$, a parameter used for tuning the widening precision~\cite{bartzis2004,silva2006}. For instance, let us consider the automata $\aut, \aut^\prime \in \fa$ recognizing the languages $\lang = \{\epsilon, a\}$ and $\lang^\prime = \{\epsilon, a, aa\}$, respectively. The result of the application of the widening $\widfa$, with $n = 1$, is $\aut \mathbin{\widfa} \aut^\prime = \aut^\second$ s.t. $\lang(\aut^\second) = \sset{a^n}{n \in \nats}$.

\noindent\textbf{Core language and semantics.} $\;$ 
\begin{figure}[t]
	\begin{framed}
		\vbox{%
			\setlength{\grammarparsep}{3pt plus 1pt minus 1pt} 
			\setlength{\grammarindent}{4em} 
			\renewcommand{\syntleft}{}  \renewcommand{\syntright}{}
			\begin{grammar}	
				
				<$\aexp \in$ \aexps> $::=$ $x\in\ids$ ~|~ $n\in\ints$  
				~|~ $\aexp$ \tt{+} $\aexp$
				~|~ $\aexp$ \tt{-} $\aexp$
				~|~ $\aexp$ \tt{*} $\aexp$
				~|~ $\aexp$ \tt{/} $\aexp$
				\alt \length{$\sexp$}
				~|~	 \indexof{$\sexp$}{$\sexp$}
				
				<$\bexp \in$ \bexps> $::=$ $x \in\ids$ ~|~ \true ~|~ \false
				~|~ $\bexp$ \tt{\&\&} $\bexp$
				~|~ $\bexp$ \tt{||} $\bexp$
				~|~ \tt{!} $\bexp$
				\alt $\exp$ \tt{\textless} $\exp$
				~|~ $\exp$ \tt{==} $\exp$
				~|~ \contains{$\sexp_1$}{$\sexp_2$}
				
				<$\sexp \in$ \sexps> $::=$ $x\in\ids$ ~|~ $\str{\sigma}$
				~|~ \subs{$\sexp$}{$\aexp$}{$\aexp$}
				\alt \concat{$\sexp$}{$\sexp$}
				~|~ \replace{$\sexp$}{$\sexp$}{$\sexp$} $\qquad (\sigma \in \Sigma^*)$
				
				<$\exp \in$ \exps> $::=$ $\aexp$
				~|~ $\bexp$
				~|~ $\sexp$
				
				<$\stmt \in$ \stmts> $::=$ $\stmt$ {\tt ;} $\stmt$ ~|~ $\ski$ ~|~ $x$  {\tt =} $\exp$ 
				~|~ \ifc{$\bexp$}{$\stmt$}{$\stmt$} 
				\alt \while{$\bexp$}{$\stmt$}
				
				<$\prog \in \imp$> $::=$ $\stmt$ {\tt ;}
			\end{grammar}
		}%
		\vspace{-14pt}
	\end{framed}
	\caption{$\imp$ syntax}
	\label{fig:imp-syntax}
	\vskip-15pt
\end{figure}
We introduce a minimal core language $\imp$, whose syntax is reported in Fig.~\ref{fig:imp-syntax}. Such language supports the main operators over strings. In particular, $\imp$ supports arithmetic expressions ($\aexps$), Boolean expressions ($\bexps$) and string expressions ($\sexps$). Primitives values are $\val = \ints \cup \Sigma^* \cup \{\true, \false\}$, namely integers, strings and booleans. Programs states $\Mem : \ids \rightarrow \val$ map identifiers to primitives value, ranged over the meta-variables $\mem$. The concrete semantics of $\imp$ statements is captured by the function $\csem{\stmt} : \Mem \rightarrow \Mem$. The semantics is defined in a standard way, and it is reported in Appendix~\ref{sect:impstsem}. Such semantics relies on the one of expressions, that we capture, abusing notation, as $\csem{\exp} : \Mem \rightarrow \val$. While the semantics concerning arithmetic and Boolean expressions is straightforward (and not of interest of this paper), we define the part concerning strings in Fig.~\ref{imp:expressions}.

\begin{figure}[t]
	\begin{align*}
	\csem{\subs{\sexp}{\aexp}{\aexp'}}\mem &= 
	\sigma_{i}\dots \sigma_{j} \qquad \mbox{if } i \leq j < |\sigma| \\
	\csem{\length{\sexp}}\mem &= |\sigma|\\
	\csem{\indexof{\sexp}{\sexp'}}\mem &= \begin{cases}
	\min\sset{i}{\sigma_i\dots\sigma_j = \sigma'} & \mbox{if }\exists i,j\in\nats\st \sigma_i\dots\sigma_j = \sigma' \\
	-1 & \mbox{otherwise}
	\end{cases}\\
	\csem{\replace{\sexp}{\sexp'}{\sexp''}}\mem &= \begin{cases}
	\sigma[\sigma' / \sigma'']& \mbox{if } \sigma' \sub \sigma\\
	\sigma & \mbox{otherwise}
	\end{cases}\\
	\csem{\concat{\sexp}{\sexp'}}\mem &= \sigma\cdot\sigma'\\
	\csem{\contains{\sexp}{\sexp'}}\mem &= \begin{cases}
	\true & \mbox{if } \exists i,j\in\nats\st\sigma_i\dots\sigma_j = \sigma'\\
	\false & \mbox{otherwise}
	\end{cases}
	\end{align*}
	\caption{Concrete semantics of $\imp$ string expressions}
	\label{imp:expressions}
\end{figure}

\section{The $\tarsis$ abstract domain}\label{sect:domain}

In this section, we recast the original finite state abstract domain working over an alphabet of character $\Sigma$, reported in Sect.~\ref{sec:bg}, to an augmented abstract domain based on finite state automata over an alphabet of strings.

\subsection{Abstract domain and widening}\label{sect:domandwid}

The key idea of $\tarsis$ is to adopt the same abstract domain, changing the alphabet on which finite state automata are defined to a set of strings, namely $\Sigma^*$. Clearly, the main concern here is that $\Sigma^*$ is infinite and this would not permit us to adopt the finite state automata model, that requires the alphabet to be finite. Thus, in order to solve this problem, we make such abstract domain \textit{parametric} to the program we aim to analyze and in particular to its strings. Given an $\imp$ program $\prog$, we denote by $\Sigma^*_\prog$ any substring of strings appearing in $\prog$\footnote{The set $\Sigma^*_\prog$  can be easily computed collecting the constant strings in $\prog$ by visiting its abstract syntax tree and then computing their substrings.}. The alphabet $\Sigma^*_\prog$ contains any possible string that can be computed by the program  $\prog$, \textit{delimiting} the space of string properties we aim to check on
$\prog$.

At this point, we can instantiate the automata-based framework proposed in~\cite{arceri2019-fa} with the new alphabet as
$$
\latticehfa
$$
The alphabet on which finite state automata are defined is  $\alphabet{\prog} \defn \Sigma^*_\prog \cup \{\ctop\}$, where $\ctop$ is a special symbol that we intend as \textit{"any possible string"}. Let $\HFA$ be the set of any deterministic finite state automaton over the alphabet $\alphabet{\prog}$. Thus, $\hfa$ is the quotient set of $\HFA$ \wrt the equivalence relation induced by language equality. $\leqhfa$ is the partial order induced by language inclusion, $\lubhfa$ and $\glbhfa$ are the lub and the glb corresponding to the union and the intersection automata operations, respectively. The bottom element is $\minimize(\varnothing)$, corresponding to the automaton recognizing the empty language and the maximum is $\minimize(\alphabet{\prog}^*)$, namely the automaton recognizing any string over~$\alphabet{\prog}$.
%
%
%
%

Like in the standard finite state automata domain $\fa$, also $\hfa$ is not a complete lattice and, consequently, it does not form a Galois Connection with the string concrete domain $\wp(\Sigma^*)$. This comes from the non-existence, in general, of the best abstraction of a strings set in $\hfa$ (e.g., a context-free language has no best abstract element in $\hfa$ approximating it). Nevertheless, this is not a concern since weaker forms of abstract interpretation are still possible~\cite{cc92} still guaranteeing soundness relations between concrete and abstract elements (e.g., polyhedra~\cite{ch78}). In particular, also without having the best abstraction, we can still ensuring soundness comparing the concretizations of our abstract elements (cf. Sect. 8 of~\cite{cc92}). Hence, we define the concretization function $\gammahfa : \hfa \rightarrow \wp(\Sigma^*)$  as
$
\gammahfa(\aut) \defn \bigcup_{\sigma \in \lang(\aut)} \fla(\sigma)
$,
where $\fla$ converts a string over $\alphabet{\prog}$ into a set of strings over $\Sigma^*$. For instance $\fla(a \; \ctop\ctop \; bb \; c) = \sset{a\sigma bbc}{\sigma \in\Sigma^*}$.

\paragraph*{Widening.} Similarly to the standard automata domain $\fa$, also $\hfa$ does not satisfy ACC, meaning that fix-point computations over $\hfa$ may not converge in a finite time. Hence, we need to equip $\hfa$ with a widening operator to ensure the convergence of the analysis. We define the widening operator $\widhfa{n} : \hfa \times \hfa \rightarrow \hfa$, parametric in $n \in \nats$, taking two automata as input and returning an over-approximation of the least upper bounds between them, as required by widening definition. We rely on the standard automata widening reported in Sect.~\ref{sec:bg}, that, informally speaking, can be seen as a \textit{subset construction} algorithm~\cite{davis1994} up to languages of strings of length $n$. 
In order to explain the widening $\widhfa{n}$, 
%
%
consider the following function manipulating strings.\footnote{For the sake of readability, in the program examples presented in this paper \lstinline{+} operation between strings corresponds to the string concatenation.}

\begin{CenteredBox}
	\begin{lstlisting}[escapeinside={(*}{*)}]
	function f(v) {
	res = "";
	while ((*?*)) 
	res = res + "id = " +  v;
	return res;
	}
	\end{lstlisting}
\end{CenteredBox}

The function {\tt f} takes as input parameter {\tt v} and returns variable {\tt res}. Let us suppose that {\tt v} is a statically unknown string, corresponding to the automaton recognizing $\ctop$ (i.e., $\minimize(\{\ctop\})$). The result of the function {\tt f} is a string of the form $\mathtt{id = } \ctop$, repeated zero or more times. 
Since the {\tt while} guard is unknown, the number of iterations is statically unknown, and in turn, also the number of performed concatenations inside the loop body.
The goal here is to over-approximate the value returned by the function {\tt f}, i.e., the value of {\tt res} at the~end~of the function.

\begin{figure}[t]
	\begin{subfigure}[b]{0.41\textwidth}
		\centering
		\begin{tikzpicture}[->,>=stealth',shorten >=1pt,auto,node distance=1.9cm, semithick]
		\node[initial,state,scale=\nodesize, initial text =, accepting] (0)                    {$q_0$};
		\node[state,scale=\nodesize]    (1) [right of=0] {$q_1$};
		\node[state,scale=\nodesize, accepting]    (2) [right of=1] {$q_2$};

		\path[->] (0) edge node {{\tt id = }} (1);		
		\path[->] (1) edge node {$\ctop$} (2);	
		\end{tikzpicture}
		\caption{Value of {\tt res} ($\aut$) at the beginning of the 2nd iteration of the loop}
		\label{fig:1-it}
	\end{subfigure}
	\begin{subfigure}[b]{0.59\textwidth}
		\centering
		\begin{tikzpicture}[->,>=stealth',shorten >=1pt,auto,node distance=1.9cm, semithick]
		\node[initial,state,scale=\nodesize, initial text =] (0)                    {$q_0$};
		\node[state,scale=\nodesize]    (1) [right of=0] {$q_1$};
		\node[state,scale=\nodesize]    (2) [right of=1] {$q_2$};
		
		\node[state,scale=\nodesize]    (3) [right of=2] {$q_3$};
		\node[state,scale=\nodesize, accepting]    (4) [right of=3] {$q_4$};

		\path[->] (0) edge node {{\tt id = }} (1);		
		\path[->] (1) edge node {$\ctop$} (2);	
		\path[->] (2) edge node {{\tt id = }} (3);	
		\path[->] (3) edge node {$\ctop$} (4);	
		\end{tikzpicture}
		\caption{Value of {\tt res} ($\aut'$) at the end of the 2nd iteration of the loop}
		\label{fig:2-it}
	\end{subfigure}
	\begin{subfigure}[b]{0.5\textwidth}
		\centering
		\begin{tikzpicture}[->,>=stealth',shorten >=1pt,auto,node distance=1.9cm, semithick]
		\node[initial,state,scale=\nodesize, initial text =, accepting] (0)                    {$q_0,q4$};
		\node[state,scale=\nodesize]    (1) [right of=0] {$q_1$};
		\node[state,scale=\nodesize, accepting]    (2) [right of=1] {$q_2$};
		\node[state,scale=\nodesize]    (3) [right of=2] {$q_3$};

		\path[->] (0) edge node {{\tt id = }} (1);		
		\path[->] (1) edge node {$\ctop$} (2);	
		\path[->] (2) edge node {{\tt id = }} (3);	
		\path[->] (3) edge[bend right=40] node[swap] {$\ctop$} (0);	
		\end{tikzpicture}
		\caption{The result of $\aut \widhfa{2} \aut'$}
		\label{fig:3-it}
	\end{subfigure}
	\begin{subfigure}[b]{0.5\textwidth}
		\centering
		\begin{tikzpicture}[->,>=stealth',shorten >=1pt,auto,node distance=2.5cm, semithick]
		\node[initial,accepting, state,scale=\nodesize, initial text =] (0)                    {$q_0$};
		\node[state,scale=\nodesize]    (1) [right of=0] {$q_1$};

		\path[->] (0) edge node {{\tt id = }} (1);		
		\path[->] (1) edge[bend right=50] node[swap] {$\ctop$} (0);	
		\end{tikzpicture}
		\caption{Minimized version of $\aut \widhfa{2} \aut'$}
		\label{fig:4-it}
	\end{subfigure}
	\caption{Example of widening application}
	\label{fig:widening}
	\vskip-20pt
\end{figure}

Let $\aut$, reported in Fig.~\ref{fig:1-it}, be the  automaton abstracting the value of {\tt res} before starting the second iteration of the loop, and let $\aut'$, reported in  Fig.~\ref{fig:2-it} be the automaton abstracting the value of {\tt res} at the end of the second iteration. At this point, we want to apply the widening operator $\widhfa{n}$, between $\aut$ and $\aut'$, working as follows. We first compute $\aut \lubhfa \aut'$ (corresponding to the automaton reported in Fig.~\ref{fig:2-it} except that also $q_0$ and $q_2$ are final states). On this automaton, we merge any state that recognizes the same strings of length $n$, with $n \in \nats$. In our example, let $n$ be $2$. The resulting automaton is reported in Fig.~\ref{fig:3-it}, where $q_0$ and $q_4$ are put together, the other states are left as singletons since they cannot be merged with no other state. Fig.~\ref{fig:4-it} depicts the minimized version of Fig.~\ref{fig:3-it}.

The widening $\widhfa{n}$ has been proved to meet the widening requirements (i.e., over-approximation of the least upper bounds and convergence on infinite ascending chains) in~\cite{silva2006}. The parameter $n$, tuning the widening precision, is arbitrary and can be chosen by the user. As highlighted in~\cite{arceri2019-fa}, the higher $n$ is, the more the corresponding widening operator is precise in over-approximating lubs of infinite ascending chains (i.e., in fix-point computations).

A classical improvement on widening-based fix-point computations is to integrate a threshold~\cite{cortesi2011}, namely widening is applied to over-approximate lubs  when a certain threshold (usually over some property of abstract values) is overcome. In fix-point computations, we decide to apply the previously defined widening $\widhfa{n}$ only when the number of the states of the lubbed automata overcomes the threshold $\tau \in \nats$. This permits us to postpone the widening application, getting more precise abstractions when the automata sizes do not overcome the threshold. At the moment, the threshold $\tau$ is not automatically inferred, since it surely requires further investigations.

\subsection{String abstract semantics of $\imp$}\label{sect:impabssem}

In this section, we define the abstract semantics of the string operators defined in Sect.~\ref{sec:bg} over the new string domain $\hfa$. Since $\imp$ supports strings, integers and booleans values, we need a way to merge the corresponding abstract domains. In particular, we abstract integers with the well-known interval abstract domain~\cite{cc77} defined as 
$
\intervals \defn \sset{[a,b]}{a,b \in \ints \cup \{-\infty,+\infty\}, a \leq b} \cup \{\botintervals\}
$
\noindent
and Booleans with $\bools \defn \wp(\{\true, \false\})$. As usual, we denote by $\lubintervals$ and $\lubbools$ the lubs between intervals and Booleans, respectively. In particular, we merge such abstract domains in $\aval$ by the coalesced sum abstract domain~\cite{arceri2017} as
$$
\aval \defn \hfa \oplus \intervals \oplus \bools
$$
Informally, the coalesced sum abstract domain introduces a new bottom and top element, and it \textit{coalesces} the bottom elements of the involved domains. 

The program state is represented through abstract program memories $\aMem : \ids  \rightarrow \aval$ from identifiers to abstract values. The abstract semantics is captured by the function $\asem{\stmt} : \aMem \rightarrow \aMem$, relying on the abstract semantics of expression defined by, abusing notation, $\asem{\exp} : \aMem \rightarrow \aval$. We focus on the abstract semantics of string operations\footnote{Since the abstract semantics of {\tt concat} does not add any further important technical detail to the paper, it is reported in Appendix~\ref{sect:otherops}.}, while the semantics of the other expressions is standard and does not involve strings.

\noindent\textbf{Length} $\;$ Given $\aut \in \hfa$, the abstract semantics of {\tt length} returns an interval $\left[c_1, c_2\right]$ such that $\forall \sigma \in \lang(\aut) \st c_1 \le |\sigma| \le c_2$. We recast the original idea of the abstract semantics of {\tt length} over standard finite state automata. Let $\sexp \in \sexps$, supposing that $\asem{\sexp}\amem = \aut \in \hfa$. The {\tt length} abstract~semantics~is:
$$
\asem{\length{\sexp}}\amem \defn
\begin{cases}
[|\minpath{\aut}|, +\infty] & \mbox{if } \hascycle{\aut} \lor \readstop{\aut}\\
[|\minpath{\aut}|, |\maxpath{\aut}|] & \mbox{otherwise}
\end{cases}
$$
where $\readstop{\aut}\Leftrightarrow\exists q,q'\in Q\st(q, \ctop, q') \in \delta$. Note that, when evaluating the length of the minimum path, $\ctop$ is considered to have a length of $0$. For instance, consider the automaton $\aut$ reported in Fig.~\ref{fig:length1}. The minimum path of $\aut$ is $(q_0, aa, q_1), (q_1, \ctop, q_2), (q_0, bb, q_4)$ and its length is 4. Since a transition labeled with $\ctop$ is in $\aut$ (and its length cannot be statically determined), the abstract {\tt length} of $\aut$ is $[4, +\infty]$. Consider the automaton $\aut'$ reported in Fig.~\ref{fig:length2}. In this case, $\aut'$ has no cycles and has no transitions labeled with $\ctop$ and the length of any string recognized by $\aut'$ can be determined. The length of the minimum path of $\aut'$ is 3 (below path of $\aut'$), the length of the maximum path of $\aut'$ is 7 (above path of $\aut'$) and consequently the abstract {\tt length} of $\aut'$ is $[4,7]$.

\begin{figure}[t]
	\begin{subfigure}[b]{0.5\textwidth}
		\centering
		\begin{tikzpicture}[->,>=stealth',shorten >=1pt,auto,node distance=1.7cm, semithick]
		\node[initial,state,scale=\nodesize, initial text =] (0) {$q_0$};
		\node[state,scale=\nodesize] (1) [above right of=0] {$q_1$};
		\node[state,scale=\nodesize] (2) [right of=1] {$q_2$};
		\node[state,scale=\nodesize] (3) [right of=0] {$q_3$};
		\node[state,scale=\nodesize, accepting] (4) [right of=3] {$q_4$};
		
		\path[->] (0) edge node {$aa$} (1);		
		\path[->] (1) edge node {$\ctop$} (2);		
		\path[->] (2) edge node {$bb$} (4);		
		\path[->] (0) edge node[swap] {$bbb$} (3);		
		\path[->] (3) edge node[swap] {$bbb$} (4);
		\end{tikzpicture}
		\caption{}
		\label{fig:length1}
	\end{subfigure}
	~
	\begin{subfigure}[b]{0.5\textwidth}
		\centering
		\begin{tikzpicture}[->,>=stealth',shorten >=1pt,auto,node distance=1.5cm, semithick]
		\node[initial,state,scale=\nodesize, initial text =] (0) {$q_0$};
		\node[state,scale=\nodesize] (1) [right of=0] {$q_1$};
		\node[state,scale=\nodesize] (2) [right of=1] {$q_2$};
		\node[state,scale=\nodesize] (4) [below of=1] {$q_4$};
		\node[state,scale=\nodesize] (3) [right of=4] {$q_3$};
		\node[state,scale=\nodesize, accepting] (5) [right of=2] {$q_5$};
		
		\path[->] (0) edge node {$aa$} (1);		
		\path[->] (1) edge node {$bbb$} (2);		
		\path[->] (2) edge node {$cc$} (5);		
		\path[->] (0) edge node[swap] {$a$} (4);		
		\path[->] (4) edge node {$b$} (3);		
		\path[->] (3) edge node[swap] {$c$} (5);
		\end{tikzpicture}
		\caption{}
		\label{fig:length2}
	\end{subfigure}
	\caption{(a) $\aut$ s.t. $\lang(\aut) = \{bbb\;bbb, aa \;\ctop\;bb\}$, (b) $\aut'$ s.t. $\lang(\aut') = \{a\;b\;c, aa\;bbb\;cc\}$}
	\label{fig:length}
\end{figure}

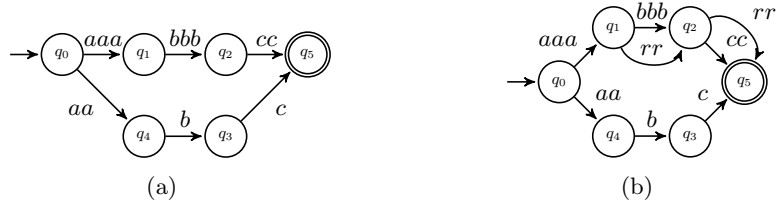
\begin{figure}[t]
	\begin{subfigure}[b]{0.5\textwidth}
		\centering
		\begin{tikzpicture}[->,>=stealth',shorten >=1pt,auto,node distance=1.6cm, semithick]
		\node[initial,state,scale=\nodesize, initial text =] (0) {$q_0$};
		\node[state,scale=\nodesize] (1) [right of=0] {$q_1$};
		\node[state,scale=\nodesize] (2) [right of=1] {$q_2$};
		\node[state,scale=\nodesize] (4) [below of=1] {$q_4$};
		\node[state,scale=\nodesize] (3) [right of=4] {$q_3$};
		\node[state,scale=\nodesize, accepting] (5) [right of=2] {$q_5$};
		
		\path[->] (0) edge node {$aaa$} (1);		
		\path[->] (1) edge node {$bbb$} (2);		
		\path[->] (2) edge node {$cc$} (5);		
		\path[->] (0) edge node[swap] {$aa$} (4);		
		\path[->] (4) edge node {$b$} (3);		
		\path[->] (3) edge node[swap] {$c$} (5);
		\end{tikzpicture}
		\caption{}
		\label{fig:replace1}
	\end{subfigure}
	~
	\begin{subfigure}[b]{0.5\textwidth}
		\centering
		\begin{tikzpicture}[->,>=stealth',shorten >=1pt,auto,node distance=1.5cm, semithick]
		\node[initial,state,scale=\nodesize, initial text =] (0) {$q_0$};
		\node[state,scale=\nodesize] (1) [above right of=0] {$q_1$};
		\node[state,scale=\nodesize] (2) [right of=1] {$q_2$};
		\node[state,scale=\nodesize] (4) [below right of=0] {$q_4$};
		\node[state,scale=\nodesize] (3) [right of=4] {$q_3$};
		\node[state,scale=\nodesize, accepting] (5) [above right of=3] {$q_5$};
		
		\path[->] (0) edge node {$aaa$} (1);		
		\path[->] (1) edge node {$bbb$} (2);		
		\path[->] (2) edge node {$cc$} (5);		
		\path[->] (0) edge node {$aa$} (4);		
		\path[->] (4) edge node {$b$} (3);		
		\path[->] (3) edge node {$c$} (5);

		\path[->] (1) edge[bend right=69] node {$rr$} (2);		
		\path[->] (2) edge[bend left=71] node {$rr$} (5);
		\end{tikzpicture}
		\caption{}
		\label{fig:replace2}
	\end{subfigure}
	\caption{Example of may-replacement}
	\label{fig:replace}
\end{figure}

\noindent\textbf{Contains} $\;$ Given $\aut, \aut' \in \hfa$, the abstract semantics of {\tt contains} should return $\true$ 
if any string of $\aut'$ is contained into any string of $\aut$, $\false$ if any string of $\aut'$ is not surely contained in any string of $\aut$ and $\{\true, \false\}$ in the other cases. For instance, consider the automaton $\aut$ depicted in Fig.~\ref{fig:replace1} and suppose to check if it contains the automaton $\aut'$ recognizing the language $\{aa,a\}$. The automaton $\aut'$ is a \textit{single-path automaton}~\cite{arceri2019}, meaning that any string of $\aut'$ is a prefix of its longest string. In this case, the containment of the longest string (on each automaton path) implies the containment of the others, such as in our example, namely it is enough to check that the longest string of $\aut'$ is contained into $\aut$. Note that, a single-path automaton cannot read the symbol $\ctop$. We rely on the predicate $\mathsf{singlePath}(\aut)$ when $\aut$ is a non-cyclic single-path automaton and we denote by $\sigma_{\mathsf{sp}}$ its longest string.
Let $\sexp, \sexp' \in \sexps$, supposing that $\asem{\sexp}\amem = \aut \in \hfa$, $\asem{\sexp'}\amem = \aut' \in \hfa$. The {\tt contains} abstract semantics~is:
$$
\asem{\contains{\sexp}{\sexp'}}\amem \defn 
\begin{cases}
\false & \mbox{if } \aut' \glbhfa \mathsf{FA}(\aut) = \minimize(\varnothing) \\
\true & \mbox{if } \neg\hascycle{\aut} \land \mathsf{singlePath}(\aut')\\
& \land \forall \pi \in \paths{\aut}\st \sigma_{\mathsf{sp}} \sub \sigma_{\pi} \\
\{\true, \false\} & \mbox{otherwise}
\end{cases}
$$

In the first case, we denote by $\mathsf{FA}(\aut)$ the factor automaton of $\aut$, i.e.,  the automaton recognizing any substring of $\aut$. In particular, if $\aut$ does not share any substring of $\aut'$, the abstract semantics safely returns $\false$ (checking the emptiness of the greatest lower bound between $\mathsf{FA}(\aut)$ and $\aut'$). Then, if $\aut'$ is a single path automaton and $\aut$ is not cyclic, the abstract semantics returns $\true$ if any path of $\aut$ reads the longest string of $\aut'$. Otherwise,  $\{\true, \false\}$ is returned.


\noindent\textbf{IndexOf} $\;$ Given $\aut, \aut' \in \hfa$, the {\tt indexOf} abstract semantics returns an interval of the first positions of the strings of $\lang(\aut')$ inside strings of $\lang(\aut)$, recalling that when there exists a string of $\lang(\aut')$ that is not a substring of at least one string of $\lang(\aut')$, the resulting interval must take into account -1 as well. 
Let $\sexp, \sexp' \in \sexps$ and suppose $\asem{\sexp}\amem = \aut$ and $\asem{\sexp'}\amem = \aut'$. The abstract semantics of {\tt indexOf} is defined as:
$$
\asem{\indexof{\sexp}{\sexp'}}\amem \defn 
\begin{cases}
[-1, +\infty] & \mbox{if } \hascycle{\aut} \lor \hascycle{\aut'} \lor \readstop{\aut'}\\
\left[-1, -1\right] & \mbox{if } \forall \sigma' \in \lang(\aut') \; \nexists \sigma \in \lang(\aut) \st \sigma' \sub \sigma\\
\bigsqcup^{\intervals}\limits_{\sigma \in \lang(\aut')} \mathsf{IO}(\aut, \sigma) & \mbox{otherwise}\\
\end{cases}
$$
If one of the automata have cycles or the automaton abstracting strings we aim to search for ($\aut'$) has a $\ctop$-transition, we return $[-1, +\infty]$. 
Moreover, if none of the strings recognized by $\aut'$ is contained in a string recognized by $\aut$, we can safely return the  precise interval $\left[-1, -1\right]$ since any string recognized by $\aut'$ is never a substring of a string recognized by $\aut$.\footnote{Note that this is a decidable check since $\aut$ and $\aut'$ are cycle-free, otherwise the interval $[-1, +\infty]$ would be returned in the first case.} If none of the aforementioned conditions is met, we rely on the auxiliary function $\mathsf{IO} : \hfa \times \Sigma^* \rightarrow \intervals$, that, given an automaton $\aut$ and a string $\sigma$, returns an interval corresponding to the possible first positions of $\sigma$ in strings recognized by $\aut$. Since $\aut'$ surely recognizes a finite language (i.e., has no cycles), the idea is to apply $\mathsf{IO}(\aut, \sigma)$ to each $\sigma \in \lang(\aut')$ and to return the upper bound of the resulting intervals.  
%
In particular, the function $\mathsf{IO}(\aut, \sigma)$ returns an interval $[i,j] \in \intervals$ where, $i$ and $j$ are computed as follows.

$$
i = \begin{cases}
-1 & \mbox{if } \exists \pi \in \mathsf{paths}(\aut)\st \sigma \not\sub \sigma_\pi\\
\min\limits_{\pi \in \mathsf{paths}(\aut)}\ssset{i}{\sigma \sub \sigma_\pi \land \\ \sigma_{\pi_{i}}\dots\sigma_{\pi_{i+n}} = \sigma_0\dots\sigma_{i+n}} & \mbox{otherwise}\\
\end{cases}
$$
$$
j = \begin{cases}
-1 & \mbox{if } \forall \pi \in \mathsf{paths}(\aut) \st \sigma \not\sub \sigma_\pi\\
+\infty & \mbox{if } \exists \pi \in \mathsf{paths}(\aut) \st \sigma \sub \sigma_\pi\\
& \land \exists j \in \nats \st \sigma_{\pi_{j}} = \ctop\\
\max\limits_{\pi \in \mathsf{paths}(\aut)} \ssset{i}{\sigma \sub \sigma_\pi \land \\ \sigma_{\pi_{i}}\dots\sigma_{\pi_{i+n}} = \sigma_0\dots\sigma_{i+n}} & \mbox{otherwise}\\
\end{cases}
$$

We recall that given a path $\pi$, $\sigma_{\pi_{i}}$ denotes the symbol read by the transition at the $i$-position of $\pi$ and $\sigma_\pi$ the string recognized by $\pi$.
Given $\mathsf{IO}(\aut, \sigma) = [i, j] \in\intervals$, $i$ corresponds to the minimal position where the string $\sigma$ can be found in $\aut$ for the first time, while $j$ the maximal one. Let us first focus on the computation of the minimal position. If there exists a path $\pi$ of $\aut$ s.t. $\sigma$ is not recognized by $\sigma_\pi$, then the minimal position where $\sigma$ can be found in $\aut$ does not exists and -1 is returned. Otherwise, the minimal position where $\sigma$ begins across $\pi$ is returned. Let us consider now the computation of the maximal position. If all paths of the automaton do not recognize $\sigma$, then -1 is returned. If there exists a path where $\sigma$ is recognized but the character $\ctop$ appears in the path, then $+\infty$ is returned. Otherwise, the maximal index where $\sigma$ begins across the paths of $\aut$ is returned.


\noindent\textbf{Replace} $\;$ In order to give the intuition about how the abstract semantics of {\tt replace} will work,
consider the three automata $\aut,\aut_s,\aut_r \in \hfa$. Roughly speaking, the abstract semantics of {\tt replace} substitutes strings of $\aut_s$ with strings of $\aut_r$ inside strings of $\aut$. Let us refer to $\aut_s$ as the  \textit{search automaton} and to $\aut_r$ as the \textit{replace automaton}. We need to specify two types of possible replacements, by means of the following example. Consider $\aut \in \hfa$ that is depicted in Fig.~\ref{fig:replace1} and suppose that the search automaton $\aut_s$ is the one recognizing the string $bbb$ and the replace automaton $\aut_r$ is a random automaton. In this case, the {\tt replace} abstract semantics performs a \textit{must-replace} over $\aut$, namely substituting the sub-automaton composed by $q_1$ and $q_2$ with the replace automaton $\aut_r$. Instead, let us suppose that the search automaton $\aut_r$ is the one recognizing $bbb$ or $cc$. Since it is unknown which string \textit{must} be replaced (between $bbb$ and $cc$),  the {\tt replace} abstract semantics needs to perform a \textit{may-replace}: when a string recognized by the search automaton is met inside a path of $\aut$ is leaved unaltered in the automaton and, in the same position where the string is met, the abstract {\tt replace} only extends $\aut$ with the replace automaton. An example of may replacement is reported in Fig.~\ref{fig:replace}, where $\aut$ is the one reported in Fig.~\ref{fig:replace1}, the search automaton $\aut_s$ is the one recognizing the language $\{bbb,cc\}$ and the replace automaton $\aut_r$ is the one recognizing the string $rr$. 

Before introducing the abstract semantics of {\tt replace}, we define how to replace of a string into an automaton. In particular, we define algorithm $\makeReplace$ in Alg.~\ref{alg:mkreplace}, that given $\aut \in \hfa$, a replace automaton $\aut^r$ and  $\sigma \in \Sigma^* \cup \{\ctop\}$, it returns a new automaton that is identical to $\aut$ except that $\sigma$ is replaced with $\aut^r$.

\begin{figure}[t]
	\scalebox{0.85}
	{%
		\begin{algorithm}[H]
			\KwData{$\aut^o = \tuple{Q^o, \alphabet{}, \delta^o, q^o_0, F^o}, \aut^r = \tuple{Q^r, \alphabet{}, \delta^r, q^r_0, F^r} \in \hfa, \sigma \in \Sigma^* \cup \{\ctop\}$} 
			\KwResult{$\aut \in \hfa$}
			
			$Q^{result} \leftarrow Q^o \cup Q^r$; 
			$\delta^{result} \leftarrow \delta^o \cup \delta^r$\;
			\ForEach{$\pi \in \paths{\aut^o}$}{
				\ForEach{$(q_i, \sigma_0, q_{i+1}),\dots,(q_{i+n-1}, \sigma_n, q_{i+n}) \in \pi$}{
					$\delta^{result} \leftarrow \delta^{result} \cup (q_i,\epsilon,q^r_0)$\;
					$Q^{result} \leftarrow Q^{result} \cup \sset{(q_f,\epsilon,q_{i+n})}{q_f \in F^r}$\;
					\ForEach{$k \in [i+n-1, i+1]$}{
						\uIf{$\nexists (q_k,\sigma', q) \in \delta^o : q \neq q_{k+1}$}{
							$Q^{result} \leftarrow Q^{result} \setminus \{q_k\}$\;
							$\delta^{result} \leftarrow \delta^{result} \setminus \{(q_k,\sigma', q_{k+1})\}$\;
						} 
						\textbf{else break}\;
					}
				}
			}
			\textbf{return} $\tuple{Q^{result}, \alphabet{}, \delta^{result}, q^o_0, F^o}$\;
			\caption{$\makeReplace$ algorithm}
			\label{alg:mkreplace}
		\end{algorithm}
	}%
	\vskip-20pt
\end{figure}

Alg.~\ref{alg:mkreplace} searches the given string $\sigma$ across all paths of $\aut$,
collecting the sequences of transitions that recognize the search string $\sigma$ and  extracting them from the paths of $\aut$ (lines 2-3): an $\epsilon$-transition is introduced going from the first state of the sequence to the initial state of $\aut'$, and one such transition is also introduced for each final state of $\aut'$, connecting that state with the ending state of the sequence (lines 4-5). 
Then, the list of states composing the sequence of transitions is iterated backwardly (lines 6-7), stopping at the first state that has a transition going outside of such list. All the states traversed in this way (excluding the one where the iteration stopped) are removed from the resulting automaton, with the transitions connecting them (lines 8-9), since they were needed only to recognize the string that has been replaced. Note that $\makeReplace$ corresponds to a must-replace. At this point, we are ready to define the {\tt replace} abstract semantics. In particular, if either $\aut$ or $\aut_s$ have cycles or $\aut_s$ has a $\ctop$-transition, we return $\minimize(\{\ctop\})$, namely the automaton recognizing $\ctop$. Otherwise, the {\tt replace} abstract~semantics is:
%
%
%
$$
\asem{\replace{\sexp}{\sexp_s}{\sexp_r}}\amem \defn 
\begin{cases}
%
\aut & \mbox{if } \forall \sigma_s \in \lang(\aut_s) \\ 
& \nexists \sigma \in \lang(\aut)\st \\
&  \sigma_s \sub \sigma\\
\makeReplace(\aut, \sigma_s, \aut_r) & \mbox{if } \lang(\aut_s) = \{\sigma_s\}\\
\bigsqcup\limits_{\sigma \in \lang(\aut_s)} \makeReplace(\aut, \sigma, \aut_r \lubhfa \minimize(\{\sigma\})) & \mbox{otherwise}\\
\end{cases}
$$
In the first case, if none of the strings recognized by the search automaton $\aut_s$ is contained into strings recognized by $\aut$, we can safely return the original automaton $\aut$ without any replacement.
In the special case where $\lang(\aut_s) = \{\sigma_s\}$, we return the automaton obtained by performing a replacement calling the function $\makeReplace(\aut,  \sigma_s, \aut_r)$. 
In the last case, for each each string $\sigma \in \lang(\aut_s)$, we perform a may replace of $\sigma$ with $\aut_r$: note that, this exactly corresponds to a call $\makeReplace$ where the replace automaton is $\aut_r \lubhfa \minimize(\{\sigma\})$, namely $\sigma$ is not removed.
The so far obtained automata are finally lubbed together.
%
%
%

\begin{figure}[t]
	\scalebox{0.85}
	{%
		\begin{algorithm}[H]
			\KwData{$\re$ regex over $\alphabet{}$, $i,j \in \nats$} 
			\KwResult{$\sset{(\sigma, n_1, n_2)}{\sigma \in \Sigma^*, n_1, n_2 \in \nats}$}
			
			\uIf{$j = 0 \lor \re = \varnothing$}{
				\textbf{return} $\varnothing$\;
			}
			\uElseIf{$\re = \sigma \in \Sigma^*$}{
				\lIf{$i > |\sigma|$}{
					\textbf{return} $\{ (\epsilon, i - |\sigma|, j )\}$ 
				}
				\lElseIf{$i + j > |\sigma|$}{
					\textbf{return} $\{ (\sigma_i\dots\sigma_{|\sigma|-1}, 0, j - |\sigma| + i)\}$ 
				}\lElse{
					\textbf{return} $\{ (\sigma_{i}\dots\sigma_{i+j}, 0, 0) \}$ 
				}
			}
			
			\uElseIf{$\re = \ctop$}{
				
				$\mathsf{result} \leftarrow \{(\epsilon, i - k, j) : 0 \le k \le i, k \in \nats\}$\;
				$\mathsf{result} \leftarrow \mathsf{result} \cup \sset{(\bullet^k, 0, j - k)}{0 \le k \le j, k \in \nats}$\;     
				
				\textbf{return} $\mathsf{result}$\;
			}    
			\uElseIf{$\re = \re_1\re_2$}{
				$\mathsf{result} \leftarrow \varnothing$\; 
				$\mathsf{subs}_1 \leftarrow \rsubs(\re_1, i, j)$\;
				
				\ForEach{$(\sigma_1, i_1, j_1) \in \mathsf{subs}_1$}{
					\uIf{$j_1 = 0$}{
						$\mathsf{result} \leftarrow \mathsf{result} \cup \{(\sigma_1, i_1, j_1)\}$\;
					}
					\Else{
						$\mathsf{result} \leftarrow \mathsf{result} \cup \sset{(\sigma_1 \cdot \sigma_2, i_2, j_2)}{(\sigma_2, i_2, j_2) \in \rsubs(\re_2, i_1, j_1)}$\;
					}
				}
				
				\textbf{return} $\mathsf{result}$\;
			}
			\uElseIf{$\re = \re_1 || \re_2$}{
				\textbf{return} $\rsubs(\re_1, i, j) \cup \rsubs(\re_2, i, j)$\;
			}
			
			\uElseIf{$\re = (\re_1)^*$}{
				$\mathsf{result} \leftarrow \{(\epsilon,i,j)\}$;
				$\mathsf{partial} \leftarrow \varnothing$\;
				
				\Repeat{$\mathsf{partial} \neq \varnothing$}{
					$\mathsf{result} \leftarrow \mathsf{result} \cup \mathsf{partial}$;
					$\mathsf{partial} \leftarrow \varnothing$\;

					\ForEach{$(\sigma_n, i_n, j_n) \in \mathsf{result}$}{
						\ForEach{$(\mathsf{suff}, i_s, j_s) \in \rsubs(\re_1, i_n, i_n + j_n)$}{
							\uIf{$\nexists (\sigma', k, w) \in \mathsf{result} \st \sigma' = \sigma_n \cdot \mathsf{suff} \land k = i_s \land w = j_s$}{
								$\mathsf{partial} \leftarrow \mathsf{partial} \cup \{(\sigma_n \cdot \mathsf{suff}, i_s, j_s)\}$\;
							}
							
						}
						
					}   
				}
				
				\textbf{return} $\mathsf{result}$\;
			}
			\caption{$\rsubs$ algorithm}
			\label{alg:rsubs}
		\end{algorithm}
	}%
	\vskip-20pt
\end{figure}

\noindent\textbf{Substring} $\;$ Given $\aut \in \hfa$ and two intervals $\mathsf{i}, \mathsf{j} \in \intervals$, the abstract semantics of {\tt substring} returns a new automaton $\aut'$ soundly approximating any substring from $i$ to $j$ of strings recognized by $\aut$, for any $i \in \mathsf{i}, j \in \mathsf{j}$ s.t. $i \leq j$.

Given $\aut\in \hfa$, in the definition of the {\tt substring} semantics, we rely on the corresponding regex $\re$ since the two representations are equivalent 
and regexes allow us to define a more intuitive formalization of the semantics of {\tt substring}. Let us suppose that $\asem{\sexp}\amem = \aut \in \hfa$ and let us denote by $\re$ the regex corresponding to the language recognized by $\aut$. At the moment, let us consider exact intervals representing one integer value, namely $\asem{\aexp_1}\amem = [i,i]$ and $\asem{\aexp_2}\amem = [j,j]$, with $i, j \in \ints$. In this case, the abstract semantics is defined~as:
$$
\asem{\subs{\sexp}{\aexp_1}{\aexp_2}}\amem \defn \bigsqcup \minimize(\sset{\sigma} {(\sigma, 0, 0) \in \rsubs(\re, i, j - i)}) 
$$
where $\rsubs$ takes as input a regex $\re$, two indexes $i,j \in \nats$, and computes the set of substrings from $i$ to $j$ of all the strings recognized by $\re$. In particular, $\rsubs$ is defined by Alg.~\ref{alg:rsubs} and, given a regex $\re$ and  $i, j \in \nats$, it returns a set of triples of the form $(\sigma, n_1, n_2)$, such that $\sigma$ is the \textit{partial substring} that Alg.~\ref{alg:rsubs} has computed up to now, $n_1 \in \nats$ tracks how many characters have still to be skipped before the substring can be computed
and $n_2 \in \nats$ is the number of characters Alg.~\ref{alg:rsubs} needs still to look for to successfully compute a substring. 
Hence, given $\rsubs(\re, i,j)$, the result is a set of such triples; note that given an element of the resulting set $(\sigma, n_1, n_2)$, when $n_2 = 0$ means that no more characters are needed and $\sigma$ corresponds to a proper substring of $\re$ from $i$ to $j$. Thus, from the resulting set, we can filter out the partial substrings, and retrieve only proper substrings of $\re$ from $i$ to $j$, by only considering the value of $n_2$. Full explanation about how Alg.~\ref{alg:rsubs} works can be found in Appendix~\ref{sect:otherops}.

Above, we have defined the abstract semantics of {\tt substring} when intervals are constant. When $\asem{\aexp_1}\amem = [i,j]$ and $\asem{\aexp_2}\amem = [l,k]$, with $i, j, l, k \in \ints$, the abstract semantics of {\tt substring} is
$$
\asem{\subs{\sexp}{\aexp_1}{\aexp_2}}\amem \defn \bigsqcup_{a \in [i,j], b \in [l,k], a \leq b}\bigsqcup \minimize(\sset{\sigma}{(\sigma, 0, 0) \in \rsubs(\re, a, b - a)})
$$

We do not precisely handle the cases when the intervals are unbounded (e.g., $[1, +\infty]$). These cases have been already considered in~\cite{arceri2019-fa} and treated in an ad-hoc manner and one may recast the same proposed idea in our context.
Nevertheless, when these cases are met, our analysis returns the automaton recognizing any possible substring of the input automaton, still guaranteeing~soundness.

\section{Experimental Results}\label{sec:experiments}

$\tarsis$ has been compared with five other domains, namely the prefix (\dprefix), suffix (\dsuffix), char inclusion (\dincl), bricks (\dbricks) domains (all defined in~\cite{costantini2015}), and $\fa$. 
Since the first four domains do not deal with all the operations presented in this paper (and neither with intervals, but only integers)
the comparisons presented in Sect.~\ref{sect:comparison} will focus on the precision of these operations on small examples. Then, in Sect.~\ref{sect:qual}, we tackle more complex and real world-like programs to highlight precision and performance differences of $\tarsis$
\wrt $\fa$.

All domains have been implemented in a prototype of a static analyzer for a subset of the Java language, similar to $\imp$ (Sect.~\ref{sec:bg}), plus the \code{assert} statement.
In particular, our analyzer raises a \textit{definite} alarm (\code{DA} for short) when a failing assert is met, namely when the assertion is definitely false, while it raises a \textit{possible} alarm (\code{PA} for short) when the assertion \textit{might} fail (i.e., the assertion evaluates to $\ctop_\bools$).
%
Comparisons have been performed by analyzing the code through the coalesced sum domain specified in Sect.~\ref{sect:impabssem} with trace partitioning~\cite{tracepartitioning}, plugging in the various string domains. All experiments have been performed on a HP EliteBook G6 machine, with an Intel Core i7-8565U @ 1.8GHz processor and 16 GB of RAM memory.

\begin{figure}[t] 
	\begin{subfigure}[b]{.5\textwidth}
		\centering
		\begin{lstlisting}
		void substring() {
		String res = "substring test";
		if (nondet)
		res = res + " passed";
		else
		res = res + " failed";
		result = res.substring(5, 18);
		assert (res.contains("g"));
		assert (res.contains("p"));
		assert (res.contains("f"));
		assert (res.contains("d"));
		}
		\end{lstlisting}
		\caption{Program \progname{subs}}
		\label{code:substring}
	\end{subfigure}
	\begin{subfigure}[b]{.5\textwidth}
		\centering
		\begin{lstlisting}
		void loop() {
		String value = read();
		String res = "Repeat: ";
		while (nondet) 
		res = res + value + "!";
		assert (res.contains("t")); 
		assert (res.contains("!")); 
		assert (res.contains("f"));
		}
		\end{lstlisting}
		\caption{Program \progname{loop}}
		\label{code:loop}
	\end{subfigure}
	\caption{Program samples used for domain comparison}
	\label{code:comparisons}
	\vskip-20pt
\end{figure}

\subsection{Precision of the various domains on test cases}\label{sect:comparison}

We start by considering programs \progname{subs} (Fig.~\ref{code:substring}) and \progname{loop} (Fig.~\ref{code:loop}).
\progname{subs} calls \code{substring} on the concatenation between two strings, where the first is constant and the second one is chosen in a non-deterministic way (i.e., \code{nondet} condition is statically unknown, lines 3-6). \progname{loop} builds a string by repeatedly appending a suffix, which contains a user input (i.e., an unknown string), to a constant value. Tab.~\ref{table:approx} reports the value approximation for \code{res} for each abstract domain and analyzed program as well as if the abstract domain precisely dealt with the program assertions, when the first assertion, of each program is met. For the sake of readability, $\tarsis$ and $\fa$ approximations are expressed as regexes.

\begin{table}[b]
	\setlength{\tabcolsep}{5pt}
	\centering
	\begin{tabular}{r|cc|cc}
		\textbf{Domain} & \multicolumn{2}{c}{\textbf{Program \progname{subs}}} & \multicolumn{2}{c}{\textbf{Program} \progname{loop}}\\
		\hline
		$\dprefix$ & $\code{ring test}$ & \xmark & $\code{Repeat: }$ & \xmark\\
		$\dsuffix$ & $\epsilon$ & \xmark & $\epsilon$ & \xmark \\
		$\dincl$ & $\left[\right]\left[\code{abdefgilnprstu }\right]$ & \checkmark & $\left[\code{:aepRt }\right]\left[\code{!:aepRt }\ctop\right]$& \xmark \\
		$\dbricks$ & $\left[\left\{\code{ring test fai}, \code{ring test pas}\right\}\right](1,1)$& \xmark & $\left[\left\{\ctop\right\}\right](0,+\infty)$ & \checkmark \\
		$\fa$ & $\code{ring test }(\code{pas} || \code{fai})$ & \checkmark & $\code{Repeat: }(\ctop)^*$ & \checkmark \\
		$\tarsis$ & $(\code{ring test pas} || \code{ring test fai})$ & \checkmark & $\code{Repeat: }(\ctop\code{!})^*$ & \checkmark \\
	\end{tabular}
	\vskip5pt
	\caption{Values of \code{res} at the first assert of each program}
	\label{table:approx}
	\vskip-25pt
\end{table}

When analyzing \progname{subs}, both $\dprefix$ and $\dsuffix$ lose precision  since  the string to append to \code{res} is statically unknown.
This leads, at line 7, to a partial substring of the concrete one with $\dprefix$, and to an empty string with $\dsuffix$. Instead, the \code{substring} semantics of $\dincl$ moves every character of the receiver in the set of possibly contained ones, thus the abstract value at line 7 is composed by an empty set of included characters, and a set of possibly included characters containing the ones of both strings. Finally, $\dbricks$, $\fa$ and $\tarsis$ are expressive enough to track any string produced by any concrete execution of \progname{subs}.
%

When evaluating the assertions of \progname{subs}, a \code{PA} should be raised on lines 9 and 10, since \textit{p} or \textit{f} might be in \code{res}, together with a \code{DA} alarm on line 111, since \textit{d} is surely not contained in \code{res}. No alarm should be raised on line 8 instead, since \code{g} is part of the common prefix of both branches and thus will be included in the substring. Such behavior is achieved when using $\dbricks$, $\fa$, or $\tarsis$. Since the \code{substring} semantics of $\dincl$ moves all characters to the set of possibly contained ones, \code{PA}s are raised on all four assertions. 
Since $\dsuffix$ loses all information about \code{res}, \code{PA}s are raised on lines 7-10 when using such domain. $\dprefix$ instead tracks the definite prefix of \code{res}, thus the \code{PA} at line 7 is avoided. 

When analyzing \progname{loop}, we expect to obtain no alarm at line 6 (since character {\tt t} is always contained in the resulting string value), and \code{PA} at lines 7 and 8. $\dprefix$ infers as prefix of \texttt{res} the string \code{Repeat :}, 
keeping such value for the whole analysis of the program. This allows the analyzer to prove the assertion at line 6, but it raises \code{PA}s when it checks the ones at lines 7 and 8.
Again, $\dsuffix$ loses any information about \code{res} since the lub operation occurring at line 3 cannot find a common suffix between \code{"Repeat: "} and \code{"!"}, hence  \code{PA}s are raised on lines 6-8. Since the set of possible characters contains $\ctop$, $\dincl$ can correctly state that any character might appear in the string. For this reason, two \code{PA}s are reported on lines 7 and 8, while no alarm is raised on line 6 (again, this is possible since the string used in the \code{contains} call has length 1). The alternation of $\ctop$ and \code{!} prevents $\dbricks$ normalization algorithm from merging similar bricks. This will eventually lead to overcoming the length threshold $\code{k}_L$, hence resulting in the $\left[\left\{\ctop\right\}\right](0,+\infty)$ abstract value. In such a situation, $\dbricks$ returns $\ctop_\bools$ on all \code{contains} calls, resulting in \code{PA}s on lines 6-8. The parametric widening of $\fa$ collapses the colon into $\ctop$. In $\tarsis$, since the automaton representing \code{res} grows by two states each iteration, the parametric widening defined in Sect.~\ref{sect:domandwid} can collapse all the the whole content of the loop into a 2-states loop recognizing $\ctop\code{!}$. The precise approximation of \code{res} of both domains enable the analyzer to detect that the assertion at line 6 always holds, while \code{PA}s are raised on lines 7 and 8.

In summary, $\dprefix$ and $\dsuffix$ failed to produce the expected results on both \progname{subs} and \progname{loop}, while $\dincl$ and $\dbricks$ produced exact results in one case (\progname{loop} and \progname{subs}, respectively), but not in the other. Hence, $\fa$ and $\tarsis$ were the two only domains that produced the desired behavior in these rather simple test cases.

\subsection{Evaluation on realistic code samples}
\label{sect:qual}

\begin{figure}[t]
	\begin{subfigure}[b]{.5\linewidth}
		\begin{lstlisting}
		void toString(String[] names) {
		String res="People: {";
		int i=0;
		while(i<names.length){
		res=res+names[i];
		if(i!=names.length-1)
		res=res+",";
		i=i+1;
		}
		res=res+"}";	
		assert(res.contains("People"));
		assert(res.contains(","));
		assert(res.contains("not"));
		}
		\end{lstlisting}
		\caption{Program \progname{toString}}
		\label{code:arrays}
	\end{subfigure}
	\begin{subfigure}[b]{.5\linewidth}
		\begin{lstlisting}[extendedchars=true, escapeinside={(*}{*)}]
		void count(boolean nondet) {
		String str;
		if(nondet) str="this is the thing";
		else str="the throat";
		int count=countMatches(str, "th")
		assert(count>0); 
		assert(count==0); 
		assert(count==3);
		}
		\end{lstlisting}
		\caption{Program \progname{count}}
		\label{code:matches}
	\end{subfigure}
	\caption{Programs used for assessing domain precision}
	\vskip-20pt
\end{figure}

\begin{table}[b]
	\setlength{\tabcolsep}{5pt}
	\centering
	\begin{tabular}{r|cc|cc}
		\textbf{Domain} & \multicolumn{2}{c}{\textbf{Program \progname{toString}}} & \multicolumn{2}{c}{\textbf{Program \progname{count}}}\\
		\hline
		$\dprefix$ & $\code{People: }\{$ & \xmark & $[0, +\infty]$ & \xmark \\
		$\dsuffix$ & $\epsilon$ & \xmark & $[0, +\infty]$ & \xmark \\
		$\dincl$ & $\left[\{\}\code{:Peopl }\right]\left[\{\}\code{:,Peopl }\ctop\right]$ & \xmark & $[0, +\infty]$ & \xmark \\
		$\dbricks$ & $\left[\left\{\ctop\right\}\right](0,+\infty)$& \xmark & $[0, +\infty]$ & \xmark \\
		$\fa$ & $\code{People: }\{ (\ctop)^*\ctop \}$ & \checkmark & $[2, 3]$ & \checkmark  \\
		$\tarsis$ & $\code{People: }\{\} || \code{People: }\{(\ctop\code{,})^*\ctop\}$ & \checkmark  & $[2, 3]$ & \checkmark \\
	\end{tabular}
	\vskip5pt
	\caption{Values of \code{res} and \code{count} at the first assert of the respective program}
	\label{table:approx2}
	\vskip-25pt
\end{table}


In this section, we explore two real world code samples. Method \progname{toString} (Fig.~\ref{code:arrays}) transforms an array of names that come as string values into a single string. While it resembles the code of \progname{loop} in Fig.~\ref{code:loop} (thus, results of all the analyses show the same strengths and weaknesses), now assertions check \code{contains} predicates with a multi-character string.
Method \progname{count} (Fig.~\ref{code:matches}) makes  use of \progname{countMatches} (reported in Sect.~\ref{sec:motivating}) to prove properties about its return value. Since the analyzer is not inter-procedural, we inlined \progname{countMatches}~inside \progname{count}. 
Tab.~\ref{table:approx2} reports the results of both methods (stored in \code{res} and \code{count}, respectively) evaluated by each analysis at the first assertion, as well as if the abstract domain precisely dealt with the program assertions.

As expected, when analyzing \progname{toString}, each domain showed results similar to those of \progname{loop}. In particular, we expect to obtain no alarm at line 11 (since \code{People} is surely contained in the resulting string), and two \code{PA}s at line 12 and 13. $\dprefix$, $\dsuffix$, $\dincl$ and $\dbricks$ raise \code{PA}s on all the three assert statements. $\fa$ and $\tarsis$ detect that the assertion at line 11 always holds. Thus, when using them, the analyzer raises \code{PA}s on lines 12 and 13 since: comma character is part of \code{res} if the loop is iterated at least once, and $\ctop$ might match \code{not}.

If \progname{count} (with the inlined code from \progname{countMatches}) was to be executed, \code{count} would be either $2$ or $3$ when the first assertion is reached, depending on the choice of \code{str}. Thus, no alarm should be raised at line 6, while a \code{DA} should be raised on line 7, and a \code{PA} on line 8. Since $\dprefix$, $\dsuffix$, $\dincl$ and $\dbricks$ do not define most of the operations used in the code, the analyzer does not have information about the string on which \progname{countMatches} is executed, and thus abstract \code{count} with the interval $[0..+\infty]$. Thus, \code{PA}s are raised on lines 6-8. Instead, $\fa$ and $\tarsis$ are instead able to detect that \code{sub} is present in all the possible strings represented by \code{str}. Thus, thanks to trace partitioning, the trace where the loop is skipped and \code{count} remains $0$ gets discarded. Then, when the first \code{indexOf} call happens, $[0, 0]$ is stored into \code{idx}, since all possible values of \code{str} start with \code{sub}. Since the call to \code{length} yields $[10, 17]$, all possible substrings from $[2, 2]$ (\code{idx} plus the length of \code{sub}) to $[10, 17]$ are computed (namely, \code{"e throat"}, \code{"is is th"}, \code{"is is the"}, \dots, \code{"is is the thing"}), and the resulting automaton is the one that recognizes all of them. Since the value of \code{sub} is still contained in every path of such automaton, the loop guard still holds and the second iteration is analyzed, repeating the same operations. When the loop guard is reached for the third time, the remaining substring of the shortest starting string (namely \code{"roat"}) recognized by the automaton representing \code{str} will no longer contain \code{sub}: a trace where \code{count} equals $[2, 2]$ will leave the loop. 
A further iteration is then analyzed, after which \code{sub} is no longer contained in any of the strings that \code{str} might hold. Thus, a second and final trace where \code{count} equals $[3, 3]$ will reach the assertions, and will be merged by interval lub, obtaining $[2, 3]$ as final value for \code{count}. This allows $\tarsis$ and $\fa$ to identify that the assertion at line 7 never holds, raising a \code{DA}, while the one at line 8 might not hold, raising a \code{PA}.

\subsection{Efficiency}

\begin{table}[t]
	\setlength{\tabcolsep}{5pt}
	\centering
	\begin{tabular}{r|c|c|c|c}
		\textbf{Domain} & \textbf{\progname{subs}} & \textbf{\progname{loop}} & \textbf{\progname{toString}} & \textbf{\progname{count}}\\
		\hline
		$\dprefix$ & 11 ms & 3 ms & 78 ms & 29 ms \\
		$\dsuffix$ & 10 ms & 2 ms & 92 ms & 29 ms \\
		$\dincl$ & 10 ms & 3 ms & 90 ms & 29 ms \\
		$\dbricks$ & 13 ms & 3 ms & 190 ms & 28 ms \\
		$\fa$ & 10 ms & 52013 ms & 226769 ms & 4235 ms \\
		$\tarsis$ & 34 ms & 38 ms & 299 ms & 39 ms \\
	\end{tabular}
	\vskip5pt
	\caption{Execution times of the domains on each program}
	\label{table:times}
\end{table}

The detailed analysis of two test cases, and two examples taken from real-world code underlined that $\tarsis$ and $\fa$ are the only ones able to obtain precise results on them. We now discuss the efficiency of the analyses.  Tab.~\ref{table:times} reports the execution times for all the domains on the case studies analyzed in this section. Overall, $\dprefix$, $\dsuffix$, $\dincl$, and $\dbricks$ are the fastest domains with times of execution usually below 100 msecs. Thus, if on the one hand these domains failed to prove some of the properties of interest, they are quite efficient and they might be helpful to prove simple properties. $\tarsis$ execution times are higher but still comparable with them (about about 50\% overhead on average). Instead, $\fa$ blows up on three out of the four test cases (and in particular on \progname{toString}). Hence, $\tarsis$ is the only domain that executes the analysis in a limited time while being able to prove all the properties of interest on these four case studies.

%

\section{Conclusion}\label{sec:conclusion}

In this paper we introduced $\tarsis$, an abstract domain for sound abstraction of string values. $\tarsis$ is based on finite state automata paired with their equivalent regular expression: a representation that allows precise modeling of complex string values. Experiments show that $\tarsis$ achieves great precision also on code that heavily manipulate string values, while the time needed for the analysis is comparable with the one of other simpler domains.


The analysis proposed in this paper is intra-procedural and we are currently working on extending it to an inter-procedural analysis.
Moreover, in order to further improve the performance of our analysis,  sophisticated techniques such as abstract slicing~\cite{mastroeni2010,mastroeni2017} can be integrated to keep the size of automata arising during abstract computations as low as possible, by focusing the analysis only on the string variables of interest. Finally, in this paper, we did not investigate completeness property of $\tarsis$ w.r.t. the considered operations of interest. This would ensure that no loss of information is related to $\hfa$ due to the input abstraction process~\cite{arceri2019}. Our future directions will include a deeper study about $\hfa$ completeness, and possibly the application of completion processes when incompleteness arises for a string operation~\cite{giacobazzi2000}.

\bibliographystyle{splncs04}
\bibliography{biblio}
\newpage
\appendix

\section{Concrete semantics of $\imp$ statements}
\label{sect:impstsem}

In the following, we report the concrete semantics of $\imp$ statements, where $\csem{\sexp}\mem = \sigma$, $\csem{\sexp'}\mem = \sigma'$, $\csem{\sexp''}\mem = \sigma''$,  $\csem{\aexp}\mem = i$ and $\csem{\aexp'}\mem = j$. 

\begin{align*}
\csem{x = \exp}{\mem} &= \cupdate{\mem}{x}{\csem{\exp}{\mem}} \\
\csem{\ifc{\bexp}{\stmt_1}{\stmt_2}}{\mem} &= \begin{cases}
\csem{\stmt_1}{\mem} & \mbox{if } \csem{\bexp}{\mem} = \true \\
\csem{\stmt_2}{\mem} & \mbox{if } \csem{\bexp}{\mem} = \false
\end{cases} \\
\csem{\while{\bexp}{\stmt}}{\mem} &= \csem{\ifc{\bexp}{ \stmt ; \while{\bexp}{\stmt}}{}}{\mem} \\
\csem{\bl{}}{\mem} &= \csem{\ski}{\mem} = \mem \qquad \\
\csem{\bl{\stmt}}{\mem} &= \csem{\stmt}{\mem}  \qquad \\
\csem{\stmt_1 ; \stmt_2}{\mem} &= \csem{\stmt_2}{(\csem{\stmt_1}{\mem})} \\
\end{align*}

\section{Abstract operations}\label{sect:otherops}

In this appendix we report a detailed explanation of the $\rsubs$ algorithm (Alg.~\ref{alg:rsubs}), together with the abstract semantics of {\tt concat}.

\noindent\textbf{Substring (Alg.~\ref{alg:rsubs}).} Alg.~\ref{alg:rsubs} is defined by case on the structure of the input regex $\re$. The four base cases, namely when $j = 0$, $\re = \varnothing$, $\re = \sigma \in \Sigma^*$
and $\re = \ctop$, are defined at lines 1-10. 
\textbf{(I, II) $j = 0$ or $\re = \varnothing$ (lines 1-2)} Alg.~\ref{alg:rsubs} returns the empty set since we have terminated the recursive computation of the substrings. \textbf{(III) $\re = \sigma \in \Sigma^*$ (lines 3-6)} If $i > |\sigma|$, it means that the beginning of the requested substring is after the end of this atom, hence we return a singleton set containing the empty string $\epsilon$, also updating $n_1$ with $i - |\sigma|$, tracking the consumed character before the beginning of the requested substring, while $n_2$ is $j$. If $i + j>  |\sigma|$, he substring begins in $\sigma$ but ends in subsequent regexes. In this case, we return a singleton set containing the substring of $\sigma$ from $i$ to $|\sigma| - 1$, setting $n_1$ to 0 since we reached the beginning of the substring, while we set $n_2$ to $j - |\sigma| + i$, namely the proper number of missing characters to get the substring. Finally, in the last case, the substring is fully contained in $\sigma$, hence we return the substring of $\sigma$ between $i$ and $i+j$, setting both $n_1$ and $n_2$ to 0. \textbf{(IV) $\re = \ctop$ (lines 7-10)} Since $\re$ might have any length, we need to produce a set of strings that (i) gradually consume all the missing characters before the substring can begin (line 8) and (ii) gradually consume all the characters that make up the substring by adding the unknown character $\bullet$ (line 9). By doing so, we consider all possible lengths of $\re$ that can influence the resulting set of strings.

The inductive cases $\re = \re_1\re_2$, $\re = \re_1 || \re_2$ and $\re = (\re_1)^*$ are defined at lines 11-31. \textbf{(V) $\re = \re_1\re_2$ (lines 11-20)} In this case, the algorithm must consider the fact that the desired substring can be fully found either in $\re_1$ or $\re_2$, or it could overlap them. First, we compute all the partial substrings of $\re_1$, recursively calling $\rsubs$ (line 13). For all of such partial substrings, the ones that are fully contained in $\re_1$ (namely when $j_1 = 0$) are added to the result (lines 15-17).
Concerning the remaining partial substrings, namely the ones that require other characters from $\re_2$ in order to complete the desired substring, we first compute the partial substrings of $\re_2$ where $n_1$ and $n_2$ corresponds to the ones returned by the partial substrings of $\re_1$ (line 18), and finally we add to the final result the concatenation of the partial substrings of $\re_1$ with the ones of $\re_2$. \textbf{(VI) $\re = \re_1 || \re_2$ (lines 20-21)} We return the partial substring of $\re_1$ and the ones of $\re_2$, recursively calling $\rsubs$. \textbf{(VII) $\re = (\re_1)^*$ (lines 22-31)} Since we do not have knowledge about how many times the inner regex $\re_1$ will be repeated, we construct the set of substrings through a fixpoint algorithm: we start by assuming that $\re_1$ is repeated $0$ times (line 23), generating the $\epsilon$ string with unchanged indexes $i$ and $j$. Then, at each iteration, we join all the partial results obtained until the previous iteration with with the ones generated by a further recursive call to $\rsubs$, keeping only the joined results that are new.

\noindent\textbf{Concat.} Given $\aut, \aut' \in \hfa$, the abstract semantics of {\tt concat} returns a new automaton recognizing
the language $\sset{\sigma\cdot\sigma'}{\sigma\in\lang(\aut), \sigma'\in\lang(\aut')}$, that is, the concatenation between the strings of $\lang(\aut)$ with the strings of $\lang(\aut')$. This is easily achievable relying on the standard automata concatenation~\cite{davis1994}. Let $\sexp, \sexp' \in \sexps$ and suppose that $\asem{\sexp}\amem = \tuple{Q, \alphabet{}, \delta, q_0, F} \in \hfa$, $\asem{\sexp'}\amem = \tuple{Q', \alphabet{}, \delta', q'_0, F'} \in \hfa$. Then, the abstract semantics of {\tt concat} is defined~as:
$$
\asem{\concat{\sexp}{\sexp'}}\amem \defn \minimize(\tuple{Q \cup Q', \alphabet{}, \delta \cup \delta' \cup \sset{(q_f, \epsilon, q'_0)}{q_f \in F}, q_0, F'})
$$
Following the standard automata concatenation, the abstract semantics of {\tt concat} between $\aut$ with $\aut'$,
merges the two automata (i.e., their states and transitions) and introduces an $\epsilon$-transition from each final state of $\aut$ to the initial state of $\aut'$. The initial state of the new automaton is the initial state of $\aut$, while the final states are the ones of $\aut'$.

\section{Selected proofs}\label{sect:proofs}

In this appendix, we report the soundness proofs of the string operations abstract semantics reported in Sect.~\ref{sect:impabssem}. The set of collecting primitives values is denoted by $\cval \defn \wp(\Sigma^*) \cup \wp(\ints) \cup \wp(\{\true,\false\})$. We abuse notation  denoting by $\Mem : \ids \rightarrow \cval$ the set of collecting memories, ranging over $\mem$, which associate with each identifier a collecting value; we denote  by $\csem{\exp} : \Mem \rightarrow \cval$ the collecting semantics of expressions which evaluates an expression $\exp$ and returns the set of its possible values. In this Section, we consider the collecting semantics of the string expressions as discussed in the paper, that is defined as the additive lift of the concrete semantics in Fig.~\ref{imp:expressions}.

The concretization function  $\gamma_{\aval} : \aval \rightarrow \cval$ is the coalesced sum concretization function and it is  defined as follows.
$$
\gamma_{\aval}(a) \defn
\begin{cases}
\varnothing & \mbox{if } a = \bot\\
\gamma_{\intervals}(a) & \mbox{if } a \in \intervals \\ 
\gamma_{\bools}(a)& \mbox{if } a \in \bools\\ 
\gammahfa(a)& \mbox{if } a \in \hfa \\ 
\cval & \mbox{otherwise} \\
\end{cases}
$$
where $\gamma_{\intervals} : \intervals \rightarrow \wp(\ints)$ and $\gamma_{\bools} : \bools \rightarrow \wp(\{\true, \false\})$ correspond to the concretization functions of intervals and booleans, respectively. Given this value concretization function, we can define the abstract memories concretization function $\gamma : \aMem \rightarrow \Mem$ as $\gamma(\amem) \defn \sset{x \mapsto v}{v \in \gamma(\amem)}$. In the following, we remove the subscript from $\gamma$ in order to not clutter the notation, since it is clear from the context which concretization function applies.

\vspace{2ex}

\noindent
\textbf{Substring.} Recall that the abstract semantics of {\tt substring} is defined, given $\re$ the regular expression associated with $\asem{\sexp}\amem$, $\asem{\aexp_1}\amem = [i,i]$ and $\asem{\aexp_2}\amem = [j,j]$, with $i, j \in \ints$, as 
$$
\asem{\subs{\sexp}{\aexp_1}{\aexp_2}}\amem \defn \bigsqcup \minimize(\sset{\sigma} {(\sigma, 0, 0) \in \rsubs(\re, i, j - i)}) 
$$
relying on the $\rsubs$ function working on regular expressions defined in terms of Alg.~\ref{alg:rsubs}. Hence, given $\lang$ being the language associated with $\re$, in order to prove soundness we need to prove that
$$
\csem{\subs{\lang}{\{i\}}{\{j\}}} \gamma(\amem)\subseteq \gamma(\bigsqcup \minimize(\sset{\sigma} {(\sigma, 0, 0) \in \rsubs(\re, i, j - i)})).
$$
For the sake of readability, being the inputs of {\tt substring} already explained, in the rest of the proof we omit $\gamma(\amem)$ from the collecting semantics.  The proof is done by structural induction over the structure of the regular expressions.

\vspace{1ex} 
\noindent
\textit{Base cases}
\begin{itemize}
	\item $\re = \varnothing$.	In this case, $\csem{\subs{\varnothing}{\{i\}}{\{j\}}} = \varnothing$, and $\rsubs(\varnothing, i, j - i)$ returns as result $\varnothing$ (line 2 of Alg.~\ref{alg:rsubs}) satisfying the soundness condition.
	
	\item $\re = \sigma \in\ \Sigma^*$.	Suppose that $i \leq |\sigma| < j$. Then
	$\csem{\subs{\{\sigma\}}{\{i\}}{\{j\}}} = \{\sigma_i\dots \sigma_j\}$, i.e. the substring is fully contained in $\sigma$. At lines 4-6, $\rsubs(\re, i, j - i)$ checks if the substring we want to obtain is fully contained in $\sigma$ and the partial string $(\sigma_i\dots \sigma_j, 0, 0)$ is returned. When $i > |\sigma|$,  there is not substring to be searched in $\sigma$ (i.e., $\csem{\subs{\{\sigma\}}{i}{j}} = \varnothing$), and $\rsubs(\re, i, j - i)$ returns as result the partial string $(\epsilon, i - |\sigma|, j - i)$, meaning that there is no substring to be computed in $\sigma$ but taking into account that $\sigma$ has been read ($i - |\sigma|$) and no character from $\sigma$ has been taken ($j-i$). Finally,  if $i < |\sigma|$ and $j > |\sigma|$ (line 5 of $\rsubs$), there is no substring to be searched in $\sigma$ (i.e., $\csem{\subs{\{\sigma\}}{\{i\}}{\{j\}}} = \varnothing$) but part of the substring we are looking for is in $\sigma$. Hence, $\rsubs$ returns the partial string computed by the suffix of $\sigma$ from $i$, namely $\sigma_i\dots\sigma_{|\sigma|-1}$, the position from which we need to search the remaining part of the substring (i.e., $0$) and the numbers of characters we need to still look for, namely $j-i - |\sigma_i\dots\sigma_{|\sigma|-1}|)$.
	
	\item $\re = \ctop$. 
	Remember that the language recognized by $\ctop$ is any possible string, namely $\Sigma^*$. Hence, $\csem{\subs{\Sigma^*}{\{i\}}{\{j\}}} = \sset{\sigma}{|\sigma| = j -i }$. 
	The soundness proof, in this case, can be seen as a special case of the previous one, except that the length of the strings approximated by $\ctop$ is unknown a priori. The strings corresponding to the concretization of $\ctop$ can be split in three sets based on the length $|\sigma|$ of the strings: strings s.t. $i,j \leq |\sigma|$, strings s.t. $i \geq |\sigma| $ and $i <  |\sigma| \wedge j \geq |\sigma|$. 
	The substrings of the first case are computed at lines 9 returning $(\bullet^{j-i}, 0, 0)$ whose concretization corresponds to the result of the collecting semantics. In the second case, the substring we aim to compute starts in $\ctop$ but ends outside. Indeed, the desired substrings
	are still added at line 9, i.e., $(\bullet^l, 0, j-l)$ where  $l < j-i$: the position from which the remaining part of the substring must be computed is $0$ and the number of remaining characters to be read is properly computed as $j-l$. In the last case, the substrings do not start in $\ctop$. The desired substrings
	are  added at line 8, i.e., $(\epsilon, i-l, j)$ where  $0 \leq l \leq i$: since, in this case, no desired substrings can be found in $\ctop$, $\rsubs$ returns empty strings,  just  decreasing the position from which the substring must be computed, for each possible  string expressed by $\ctop$ that is shorter than $i$.
\end{itemize}

\noindent
\textit{Inductive steps}
\begin{itemize}
	\item $\re = \re_1 || \re_2$. Let $\lang, \lang_1, \lang_2 \in \wp(\Sigma^*)$ be the languages recognized by $\re$, $\re_1$ and $\re_2$, respectively. Clearly, $\lang = \lang_1 \cup \lang_2$. In this case, it is easy to see that $\csem{\subs{\lang}{\{i\}}{\{j\}}} = \csem{\subs{\lang_1}{\{i\}}{\{j\}}} \cup \csem{\subs{\lang_2}{\{i\}}{\{j\}}}$.
	For inductive hypothesis, we have that
	$$\csem{\subs{\lang_1}{\{i\}}{\{j\}}} \subseteq \gamma(\bigsqcup \minimize(\sset{\sigma} {(\sigma, 0, 0) \in \rsubs(\re_1, i, j - i)}))$$ and  $$\csem{\subs{\lang_2}{\{i\}}{\{j\}}} \subseteq \gamma(\bigsqcup \minimize(\sset{\sigma} {(\sigma, 0, 0) \in \rsubs(\re_2, i, j - i)})).$$ The function $\rsubs$, in this case, returns $\rsubs(\re_1, i, j - i) \cup \rsubs(\re_2, i, j - i)$ at lines 20-21 and hence soundness is met, as 
	\begin{align*}
	& \csem{\subs{\lang}{\{i\}}{\{j\}}}  \\
	&  = \csem{\subs{\lang_1}{\{i\}}{\{j\}}} \cup \csem{\subs{\lang_2}{\{i\}}{\{j\}}} \\
	& \subseteq \gamma(\bigsqcup \minimize(\sset{\sigma} {(\sigma, 0, 0) \in \rsubs(\re_1, i, j - i)})) \\
	& \qquad \cup \gamma(\bigsqcup \minimize(\sset{\sigma} {(\sigma, 0, 0) \in \rsubs(\re_2, i, j - i)})) \\ 
	& = \gamma(\bigsqcup \minimize(\sset{\sigma} {(\sigma, 0, 0) \in \rsubs(\re_1 || \re_2, i, j - i)})).
	\end{align*}
	\item $\re = \re_1\re_2$. Let $\lang, \lang_1, \lang_2 \in \wp(\Sigma^*)$ be the languages recognized by $\re$, $\re_1$ and $\re_2$, respectively. Clearly, $\lang = \lang_1\cdot\lang_2$. We have three cases. Let us suppose that the substrings are fully contained in $\lang_1$, namely
	$$\csem{\subs{\lang}{\{i\}}{\{j\}}} = \csem{\subs{\lang_1}{\{i\}}{\{j\}}}$$
	Hence, for inductive hypothesis, we have that
	$$\csem{\subs{\lang_1}{\{i\}}{\{j\}}} \subseteq \gamma(\bigsqcup \minimize(\sset{\sigma} {(\sigma, 0, 0) \in \rsubs(\re_1, i, j - i)})).$$
	In particular, Alg.~\ref{alg:rsubs} computes $\rsubs(\re_1, i, j - i)$ at line 13, and at lines 15-17 it adds the proper substrings of $\re_1$ to the result (returned at line 19), satisfying soundness. The case when the substrings are fully contained in $\lang_2$ is analogous.
	
	Let us consider now the case when the substrings could be straddling $\lang_1$ and $\lang_2$, meaning that they could be straddling $\re_1$ and $\re_2$. We have already shown that the substrings fully contained in $\re_1$ and $\re_2$ are added to the final results, hence, we can focus only on the strings straddling $\re_1$ and $\re_2$. At line 13, the partial substrings of $\re_1$ are computed. For the partial substrings of the form $(\re_1, i_1, j_1)$ for which there are missing characters to complete the substring (i.e., when $j_1 \neq 0$, line 18), partial substrings of $\re_2$ are also  computed at line 18 calling $\rsubs(\re_2, i_1, j_1)$, where the position from which the remaining part of the substring must be computed is $i_1$ and $j_1$ characters must be read. At this point, the concatenation between each partial string of $\re_1$ with each partial string of $\re_2$ with the proper indexes values ($i_2$ and $j_2$ returned by $\rsubs(\re_2, i_1, j_1)$, indicating the possible missing characters to complete the substring, $j_2$, and where to start to consume characters, $i_2$) are added to the result. 
	
	\item $\re = (\re_1)^*$. The proof of this case is similar to concatenation case, since $(\re_1)^*$ can be seen as an (undefined) concatenation of the regular expression $\re_1$.
\end{itemize}

\noindent
\textbf{Length.} The collecting semantics of {\tt length} is defined as the additive lift of the concrete one reported in Fig.~\ref{imp:expressions}, namely
$$
\csem{\length{\sexp}}\mem = \sset{|\sigma|}{\sigma \in \lang} \qquad \mbox{where } \csem{\sexp}\mem = \lang \in \wp(\Sigma^*)
$$
In order to prove soundness, we need to prove that, given a string expression $\sexp \in \sexps$,
$$
\forall \amem \in \aMem \st \csem{\length{\sexp}}{\gamma(\amem)} \subseteq \gamma(\asem{\length{\sexp}}{\amem}).
$$

Let us suppose that  $\asem{\sexp}\amem = \aut \in \hfa$ and $\gamma(\aut) = \lang \in \wp(\Sigma^*)$. We split the proof in the following cases.

\begin{itemize}
	\item $\aut$ is cyclic or has a $\ctop$ transition:  
	\begin{align*}
	& \csem{\length{\sexp}}\gamma(\amem) = \sset{|\sigma|}{\sigma \in \lang} \\
	& \subseteq \gamma([\min\sset{|\sigma|}{\sigma \in \lang}, + \infty]) \\
	& = \gamma([|\minpath{\aut}|, + \infty]) \\
	& = \gamma(\asem{\length{\sexp}}\amem)
	\end{align*}

	\item $\aut$ is not cyclic and has no a $\ctop$ transition:  this means that $\lang$ is a finite language.
	\begin{align*}
	& \csem{\length{\sexp}}\gamma(\amem) = \sset{|\sigma|}{\sigma \in \lang} \\
	& \subseteq \gamma([\min\sset{|\sigma|}{\sigma \in \lang}, \max\sset{|\sigma|}{\sigma \in \lang}]) \\
	& = \gamma([|\minpath{\aut}|, |\maxpath{\aut}|]) \\
	& = \gamma(\asem{\length{\sexp}}\amem)
	\end{align*}
\end{itemize}


\noindent
\textbf{IndexOf.} The collecting semantics of {\tt indexOf} is defined as the additive lift of the concrete one reported in Fig.~\ref{imp:expressions}.
In order to prove soundness, we need to prove that, given two strings expressions $\sexp, \sexp' \in \sexps$,
$$
\forall \amem \in \aMem \st \csem{\indexof{\sexp}{\sexp'}}{\gamma(\amem)} \subseteq \gamma(\asem{\indexof{\sexp}{\sexp'}}{\amem})
$$
Let us suppose that  $\asem{\sexp}\amem = \aut$, $\gamma(\aut) = \lang$ and $\asem{\sexp'}\amem = \aut'$, $\gamma(\aut') = \lang'$, where $\aut, \aut' \in \hfa$ and $\lang, \lang' \in \wp(\Sigma^*)$.
Note that, by definition of the {\tt indexOf} concrete semantics, we have that $\csem{\indexof{\sexp}{\sexp'}}{\gamma(\amem)} \subseteq \gamma([-1, \infty])$. When $\aut$ or $\aut'$ are cyclic or $\aut'$ has a $\ctop$ transition, the abstract semantics of {\tt indexOf} returns the interval $[-1, +\infty]$, guaranteeing soundness. Hence, in the following, we focus on the other remaining cases, supposing that $\aut$ and $\aut'$ are not cyclic and $\aut'$ has no $\ctop$ transitions (meaning that $\lang'$ is a finite set of strings). 

\begin{itemize}
	\item $\csem{\indexof{\sexp}{\sexp'}}\gamma(\amem) = \{-1\}$. This means that any string in $\lang'$ is not contained into any string of $\lang$, namely $\forall \sigma'\in \lang' \; \nexists \sigma \in \lang\st \sigma' \sub \sigma$. Being  $\aut$ and $\aut'$ not cyclic, we can compute the corresponding languages and check this condition. In this case, {\tt indexOf} abstract semantics returns the interval $[-1,-1]$.
	
	\item $\csem{\indexof{\sexp}{\sexp'}}\gamma(\amem) = I \subseteq \sset{n}{n \geq 0}$. This means that every string  of $\aut$ contains any string of $\aut'$, since the result of collecting semantics of {\tt indexOf} does not contain -1. 
	In this case, we recall that the abstract semantics of {\tt indexOf} relies on the auxiliary function $\mathsf{IO}$ and it is defined as 
	$$
	\asem{\indexof{\sexp}{\sexp'}}\amem = \bigsqcup^{\intervals}\limits_{\sigma \in \lang(\aut')} \mathsf{IO}(\aut, \sigma)
	$$
	namely, for each string $\sigma'\in \lang(\aut')$ computes the interval between the minimal and the maximal position where $\sigma$ can be found in $\aut$, and finally lubs the results. Hence, it is enough to prove the correctness of the function $\mathsf{IO}$. Given $\sigma'\in \lang'$, let us denote by $I_{\sigma'} \subseteq I$ the set of positions where $\sigma'$ can be found in $\lang$ and let $m, M \in I$ be the minimal and the maximal elements of $I_{\sigma'}$.  Since $-1 \notin I$, we have that in any path of $\aut$ the string $\sigma'$ is read, and {\tt indexOf} abstract semantics successfully computes $m$ looking at each path of $\aut$. As far as the maximal position is concerned, we have two cases: $\sigma'$ is found in any path of $\aut$ and the paths 
	(i) does not read the $\ctop$ symbol, (ii) does read the symbol $\ctop$. In the first case, {\tt indexOf} successfully computes $M$, while in the second case it returns $+\infty$. Soundness of the $\mathsf{IO}$ is met since in (i) $I_{\sigma'} \subseteq \gamma([m, M])$ and in (ii) $I_{\sigma'} \subseteq \gamma([m, +\infty])$ and in turn, soundness of {\tt indexOf} abstract semantics is satisfied.
	
	\item $\csem{\indexof{\sexp}{\sexp'}}\gamma(\amem) = I \subseteq \sset{n}{n \geq -1}$. The proof is analogous to the previous case.
	
\end{itemize}

\noindent
\textbf{Contains.} The collecting semantics of {\tt contains} is defined as the additive lift of the concrete one reported in Fig.~\ref{imp:expressions}.

In order to prove soundness, we need to prove that, given two string expressions $\sexp, \sexp' \in \sexps$
$$
\forall \amem \in \aMem\st \csem{\contains{\sexp}{\sexp'}}{\gamma(\amem)} \subseteq \gamma(\asem{\contains{\sexp}{\sexp'}}{\amem})
$$

Let us suppose that  $\asem{\sexp}\amem$, $\gamma(\aut) = \lang$ and $\asem{\sexp'}\amem = \aut'$, $\gamma(\aut') = \lang'$, where $\aut,\aut' \in \hfa$ and $\lang, \lang' \in \wp(\Sigma^*)$. We split the proof in the following cases.

\begin{itemize}
	\item $\csem{\contains{\sexp}{\sexp'}}\gamma(\amem) = \{\false\}$
	$$
	\csem{\contains{\sexp}{\sexp'}}{\gamma(\amem)} = \{\false\} \Rightarrow  \forall \sigma \in \lang \; \forall \sigma' \in \lang' \st \sigma' \not\sub \sigma \\ 
	$$
	If $\aut$ and $\aut'$ has no $\ctop$ transitions, the above condition is equivalent of checking the emptiness of $\mathsf{\aut} \glbhfa \aut'$, meaning that any substring of $\aut$ does not corresponds to any string of $\aut'$. When this condition is met,  {\tt contains} abstract semantics returns $\false$. If either $\aut$ or $\aut'$ has a $\ctop$ transition, the abstract semantics returns ${\true, \false}$ and soundness is met.
	
	\item $\csem{\contains{\sexp}{\sexp'}}\gamma(\amem) = \{\true\}$

	$$
	\csem{\contains{\sexp}{\sexp'}}{\gamma(\amem)} = \{\true\} \Rightarrow  \forall \sigma \in \lang \; \forall \sigma' \in \lang' \st \sigma' \sub \sigma \\ 
	$$
	We recall that an automaton is single path when any recognized string is prefix of the longest one. If $\aut$ is not cyclic and $\aut'$ is a single path automaton, {\tt contains} abstract semantics checks that any string read by a path of $\aut$ contains the longest string of $\aut'$. Hence, any string recognized by $\aut$ contains the longest string of $\aut'$, and, being single path, also any other string of $\aut'$. If so, the abstract semantics returns $\true$. In all the other cases, the abstract semantics of {\tt contains} returns $\{\true, \false\}$, and soundness is met.
	
	\item $\csem{\contains{\sexp}{\sexp'}}\amem = \{\true, \false\}$: in this case soundness is trivially satisfied.
\end{itemize}

\end{document}